\documentclass[prd,nofootinbib,showpacs,superscriptaddress, preprint, 11pt]{revtex4-1}
\pdfoutput=1
\usepackage{amsmath,amsfonts,amssymb,graphicx,natbib,bm}
\usepackage{color}
\usepackage{slashed}    
\usepackage[colorlinks,citecolor=blue]{hyperref}
\usepackage{color}
\usepackage{url}
\usepackage{cancel}
\usepackage{slashed}
\usepackage{multirow}
\usepackage{rotating}
\usepackage{braket}
\usepackage{float}
\usepackage{subfigure}

\newcommand{\be}{\begin{eqnarray*}}
\newcommand{\ee}{\end{eqnarray*}}

\newcommand{\bee}{\begin{eqnarray}}
\newcommand{\eee}{\end{eqnarray}}
\newcommand{\beeq}{\begin{equation}}
\newcommand{\eeq}{\end{equation}}

\usepackage{xspace}

\newcommand{\ba}{\begin{array}}
\newcommand{\ea}{\end{array}}
\newcommand{\bd}{\begin{displaymath}}
\newcommand{\ed}{\end{displaymath}}
\newcommand{\besub}{\begin{subequations}}
\newcommand{\eesub}{\end{subequations}}
\newcommand{\bea}{\begin{eqnarray}}
\newcommand{\eea}{\end{eqnarray}}

\def\a{\alpha}

\def\b{\beta}
\def\g{\gamma}

\def\l{\lambda}
\def\m{\mu}

\def\q2 {q^2}

\def\bt{\begin{table}}
\def\et{\end{table}}

\begin{document}
\title{Sub-TeV Singlet Scalar Dark Matter and Electroweak Vacuum Stability with Vector Like Fermions}

\author{Debasish Borah}
\email{dborah@iitg.ac.in}
\author{Rishav Roshan}
\email{rishav.roshan@iitg.ac.in}
\author{Arunansu Sil}
\email{asil@iitg.ac.in}
\affiliation{Department of Physics, Indian Institute of Technology Guwahati, Assam-781039, India }

\begin{abstract}  
We study a scalar singlet dark matter (DM) having mass in sub-TeV regime by extending the minimal scalar 
singlet DM setup by additional vector like fermions. While the minimal scalar singlet DM satisfies the relic 
and direct detection constraints for mass beyond TeV only, presence of its portal coupling with vector like fermions 
opens up additional co-annihillation channels. These vector like fermions also help in achieving electroweak 
vacuum stability all the way up to Planck scale. We find that for one generation of vector like quarks consisting 
of a $SU(2)_L$ doublet and a singlet, scalar singlet DM with mass a few hundred GeV can indeed satisfy relic, 
direct detection and other relevant constraints while also making the electroweak vacuum absolutely stable. 
The same can be achieved by introducing vector like leptons too, but with three generations. While the model 
with vector like quarks is minimal, the three generations of vector like lepton doublet and neutral singlet can 
also give rise to light neutrino mass at one loop level.
\end{abstract}
\maketitle

\section{Introduction}
Understanding the nature of the dark matter (DM) remains an outstanding problem of present day particle 
physics. Although there have been significant evidences from astrophysics suggesting the presence of this 
non-baryonic, non-luminous and collision-less form of matter in the universe, their direct evidence is still 
awaited. Information about its abundance has been obtained from WMAP \cite{Hinshaw:2012aka} and PLANCK 
satellite experiments indicating $\Omega_{\text{DM}} h^2 = 0.120\pm 0.001$ \cite{Aghanim:2018eyx} at 
68\% CL with 
$h = \text{Hubble Parameter}/(100 \;\text{km} ~\text{s}^{-1} \text{Mpc}^{-1})$ which effectively corresponds to around $26\%$ of the present universe's energy density. Due to the lack of knowledge about its nature of 
interaction, apart from the gravitational one, a plethora of possibilities of DM model building has taken place over the years. Among them, the weakly interacting 
massive particle (WIMP) paradigm \cite{Kolb:1990vq} is the most widely studied one. In such a scenario, 
a particle DM candidate typically having electroweak scale mass and interactions, is produced 
thermally in the early universe followed by freeze-out from the bath and hence leaves a relic very close to the observed DM abundance. 

Perhaps the most economical realisation of such a WIMP scenario is the extension of the standard model (SM) by a scalar singlet field having Higgs portal interaction, odd under an unbroken $Z_2$ symmetry \cite{Silveira:1985rk, McDonald:1993ex, Burgess:2000yq} and hence explaining the stability of DM, a recent update on which can be found in \cite{Athron:2017kgt}. The scenario is tightly constrained by the relic abundance and direct detection (DD) searches. In fact, present direct detection experiments such as LUX \cite{Akerib:2016vxi}, PandaX-II \cite{Tan:2016zwf,Cui:2017nnn} and XEXON1T \cite{Aprile:2017iyp,Aprile:2018dbl} allow such a DM candidate beyond 1 TeV only (except near the SM Higgs resonance region). On the other hand, the large hadron collider (LHC) bound on Higgs invisible decay width sets constraints on such a possibility even for lighter DM mass: $m_{\rm DM} < 62.5$ GeV \cite{Athron:2017kgt, ATLAS:2020cjb}. The latest measurements by the ATLAS collaboration constrains the Higgs invisible decay branching ratio to be below $13\%$ \cite{ATLAS:2020cjb} tightly constraining the coupling of scalar DM with SM Higgs for this low mass range. Hence a significant range of DM mass in this simplest framework is presently excluded which could otherwise be an interesting region for several DM (direct and indirect) and collider experiments.

Apart from explaining the DM, such a singlet scalar extension of the SM can also be useful in keeping the electroweak (EW) vacuum absolutely stable all the way till Planck scale for $m_{\rm DM}$ above 1 TeV \cite{Garg:2017iva}. As it is known that the SM Higgs quartic coupling $\lambda_H$ becomes negative 
at an intermediate scale $\Lambda_I^{SM} \sim 10^{9-10}$ GeV (depending on the precise value of the 
top quark mass) according to the renormalisation group (RG) evolution, it indicates a possible instability of the 
Higgs vacuum. Although with 
the current measured mass of the top quark (central value) $\sim$ 173.2 GeV, the EW vacuum in the SM remains metastable \cite{Isidori:2001bm,Greenwood:2008qp,Ellis:2009tp,EliasMiro:2011aa,Alekhin:2012py,Buttazzo:2013uya, Anchordoqui:2012fq,Tang:2013bz,Salvio:2015cja,Salvio:2018rv}, a more precise measurement of top quark mass may change the conclusion. In addition, such a metastability of the EW vacuum may not be a welcome feature in the context of primordial inflation\cite{Saha:2016ozn} ( and references therein). This situation would change in presence of the scalar singlet DM as the negative fermionic contribution (primarily 
due to top quark) to the RG evolution of $\lambda_H$ can now be compensated by the additional Higgs
portal interaction of the new scalar. 

Study of such scalar singlet DM has been extended in different directions where additional portal couplings of 
the scalar are engaged on top of the usual Higgs portal interaction. One such possibility is to include DM portal couplings with new vector like leptons \cite{Toma:2013bka,Giacchino:2013bta,Giacchino:2014moa, Ibarra:2014qma,Barman:2019tuo,Barman:2019aku,Barman:2019oda} and quarks \cite{Giacchino:2015hvk,Baek:2016lnv,Baek:2017ykw,Colucci:2018vxz,Biondini:2019int}. The interesting feature that came out of these 
studies is the presence of radiative corrections that not only affects the DM annihilations significantly but also 
leads to interesting observations for DM indirect searches. Such studies are also relevant for collider searches.  
Another extension of the minimal singlet scalar DM model involves additional scalar with non-zero vacuum expectation value (VEV) as done in \cite{Ghosh:2017fmr}. The purpose of the work 
was to show that in presence of such a scalar, the low-mass window (below 500 GeV) for the DM can be reopened. The second scalar is shown to also help in making the EW vacuum stable even when large neutrino Yukawa coupling is present. 

In this work, we include additional $Z_2$ odd vector like fermions (VLF): first with vector like quarks (VLQ) and then with vector like leptons (VLL) such that the scalar DM can co-annihilate with these additional fermions. This contribution affecting the relic abundance (to some extent) gets decoupled from the ones giving rise to tree level 
spin independent DM-nucleon scattering and hence DM with mass below 1 TeV is expected to be revived. Although phenomenological implications of such a setup have already been addressed elsewhere from different perspectives \cite{Toma:2013bka,Giacchino:2013bta,Giacchino:2014moa, Ibarra:2014qma,Barman:2019tuo,Barman:2019aku,Barman:2019oda,Giacchino:2015hvk,Baek:2016lnv,Baek:2017ykw,Colucci:2018vxz,Biondini:2019int}
, here we have one more motivation and that is to ensure the stability of electroweak vacuum all the way up to the Planck scale. The idea emerges from a recent observation \cite{Gopalakrishna:2018uxn} that though we know fermions coupling to the Higgs contribute negatively to the RG running of quartic coupling $\lambda_H$ 
in general, the presence of VLQs can actually affect the $SU(3)$ gauge coupling in such a way which in turn 
be useful in keeping $\lambda_H$ positive at all scales. We show that the same idea can be extended to VLL also which affect the $SU(2)$ gauge coupling in a similar way to keep $\lambda_H$ positive at all scales, though more 
number of generations of doublet-singlet VLLs are required to include.

Motivated by the possibility of having a common framework for stable electroweak vacuum and a sub-TeV scalar singlet DM, we first consider an extension of SM by a dark sector consisting of a real scalar singlet DM and two types of VLQs (all $Z_2$ odd) transforming as doublet and singlet under $SU(2)_L$ gauge symmetry of the SM. We show the new available regions of DM parameter space with mass below a TeV allowed from all relevant constraints from cosmology and direct searches. We also show that for the same DM parameter space, the electroweak vacuum stability criteria can be satisfied. In the latter half of this paper, we study a similar model but with one vector like lepton doublet and a neutral singlet per generation, both odd under the $Z_2$ symmetry. We find that at least three such generations of leptons are required to achieve absolute electroweak vacuum stability with the help of DM-Higgs 
portal coupling (to some extent) where a sub-TeV scalar singlet DM becomes allowed from all phenomenological constraints. While this scenario requires more additional fermions compared to the VLQ model, the additional 
leptons can also take part in generating light neutrino mass at one loop level.

This paper is organised as follows. In section \ref{model}, we describe particle content of our model with VLQs and DM, the relevant interactions, particle spectrum and existing constraints. In section \ref{constraints}, we summarize the constraints applicable to this scenario. Then in section \ref{DMph}, we first discuss the DM phenomenology for the model with VLQs followed by the issues related to electroweak vacuum stability in \ref{sec:highscale}. A combined analysis and parameter space are then presented in section \ref{DM_VS}. In section \ref{sec:vll}, we 
replace the VLQs by VLLs and analyse the DM and vacuum stability in details which is inclusive of constraints applicable to the scenario, allowed parameter space and a relative comparison between the two scenarios considered in this work. Finally we conclude in section \ref{conclude}. Appendices \ref{RGE_VLQ} and \ref{RGE_VLL} are provided to summarize the beta functions of the model parameters. 
   
\section{The Model with VLQ}
\label{model}

\begin{table}[]
\begin{center}
\vskip 0.5 cm
\begin{tabular}{|c|c|c|c|c|}
\hline
    Particle & $SU(3)_c$ &  $SU(2)_L$ & $U(1)_Y$  & $Z_2$          \\
    \hline
    $Q_L$ & 3 & 2 & $\frac{1}{6}$ & $+1$ \\
    $u_R$ & 3 & 1 & $\frac{2}{3}$ & $+1$ \\
    $d_R$ & 3 & 1 & $-\frac{1}{3}$ & $+1$ \\
    $l_L$ & 1 & 2 & $-\frac{1}{2}$ & $+1$ \\
    $e_R$ & 1 & 1 & $-1$ & $+1$ \\
$H$           &1   &  2     &  $\frac{1}{2}$       &   +1  \\
\hline
$S$           &1   &  1     &  0                   &   -1  \\
$\mathcal{F}$ & 3  &   2    &   $\frac{1}{6}$                    &   -1 \\
$f^{\prime}$           & 3  &   1     &  $\frac{2}{3}$                                      &   -1\\
\hline
\end{tabular}
\end{center}  
\caption{Particles and their charges under different symmetries.}
\label{t1}
\end{table}
 
As stated in the introduction, we extend the SM fermion-fields content by two types of vector-like quarks (being 
triplet under $SU(3)_c$): a $SU(2)_L$ doublet $\mathcal{F}^T = (\mathcal{F}^{\prime}_1~~\mathcal{F}_2)$, 
and a singlet $f^{\prime}$ VLQs. In addition, a SM singlet real scalar field $S$ is included which would play the role of DM. These beyond the standard model (BSM) fields are odd under the $Z_2$ symmetry while all 
SM fields are even under it. This unbroken $Z_2$ symmetry guarantees the stability of the DM, as usual. The fermion and scalar content of the model inclusive of the SM ones and their respective charges are shown in 
table \ref{t1}. The Lagrangian terms, invariant 
under the symmetries considered, involving Yukawa interactions between the BSM or dark sector fields of 
our framework and the SM fields can now be written as
\bea
-\mathcal{L}^{\rm VLQ} = M_{\mathcal{F}} \bar{\mathcal{F}} \mathcal{F} + M_{f} \bar{f^{\prime}} f^{\prime} + y \bar{\mathcal{F}} \tilde{H}f^{\prime} + \a_{1} \bar{\mathcal{F}_R} S Q_L +\a_{2} \bar{f^{\prime}_L} S~ u_R +  h.c..
\label{lag} 
\eea
Here $L,R$ denote the left and right-handed projections respectively. Note that the choice of hypercharges 
of the VLQs ($Y_F, Y_{f'}$) are guided by the consideration that they are of SM-like. For the sake of 
minimality, we choose the singlet VLQ to be of up type only, considering it to be down type though would not 
have significant impact on our results and conclusion. The bare masses of $\mathcal{F}$ and $f'$ are 
indicated by $M_{\mathcal{F}}$ and $M_{f}$ respectively. 

Turning into the scalar part of the Lagrangian, the most general renormalisable scalar potential of our model, $V(H,S)$ can be divided into three parts: (i) $V_H$: sole contribution of the SM Higgs $H$,  (ii) $V_S$ : individual contribution of the scalar singlet $S$ and (iii) $V_{\rm int}$: interaction among $H$ and $S$ fields, as 
\begin{eqnarray}
V(H,S) &=& V_H + V_S + V_{\rm int},
\label{p1}
\end{eqnarray}
where
\bea
V_{H} = -\mu_H^2 H^\dag H + \lambda_H (H^\dag H)^2; ~~V_{S} &=\frac{1}{2} M_S^2 S^2+ \frac{\lambda_S}{4!} S^4, \\
~{\rm{and}}~V_{\rm{int}} =  \frac{\lambda_{HS}}{2}(H^\dag H)S^2. &
\label{p1int}
\eea 
Once the electroweak symmetry breaking (EWSB) takes place with $v=246$ GeV as the SM Higgs VEV, masses of the two physical states, the SM Higgs boson $h$ and $S$, are found 
to be
\bea
m_h^2&=2 \lambda_H v^2, ~{\rm{and}} ~~m_{S}^{2}&= M_S^2+ \frac{\lambda_{HS}}{2}v^2.\ \, 
\label{sclrmass}   
\eea 

Note that the EWSB gives rise to a mixing among the up-type VLQs of the framework. In the basis $(\mathcal{F}_{1}^{\prime}, f^{\prime})$, the mass-matrix is given by, 
\bea
\mathcal{M}_{VLQ} = \begin{pmatrix}
    M_{\mathcal{F}} & \frac{y~v}{\sqrt{2}} \\
   \frac{ y~v}{\sqrt{2}} & M_f 
  \end{pmatrix}.
\eea
Diagonalizing the above matrix, we get the mass eigenvalues as 
\bea
m_{2, 1}&=&\frac{1}{2}\bigg{[}M_{\mathcal{F}}+M_f\pm\sqrt{(M_{\mathcal{F}}+M_f)^2-4(M_{\mathcal{F}}M_f-\frac{y^2v^2}{2})}~\bigg{]}, 
\label{vlqmass}   
\eea 
corresponding to the mass eigenstates $f$ and $\mathcal{F}_1$. We follow the hierarchy $m_1 < m_2$. 
These mass eigenstates are related to the flavor eigenstates $\mathcal{F}_{1}^{\prime}, f^{\prime}$ via the mixing angle $\theta$ as 
\bea
\begin{pmatrix}
    \mathcal{F}_1^{\prime} \\
    f^{\prime}
  \end{pmatrix} = \begin{pmatrix}
    c_\theta & s_\theta \\
    -s_\theta & c_\theta 
  \end{pmatrix}
  \begin{pmatrix}
    \mathcal{F}_1 \\
    f 
  \end{pmatrix},
\eea
where 
\bea
\text{tan} 2\theta &=& \frac{\sqrt{2}~y~v}{ M_f-M_\mathcal{F}}.
\label{tan2theta}
\eea
In the subsequent part of the analysis, we choose the physical masses $m_1$, $m_2$ and mixing angle $\theta$ as independent variables. Obviously, the other model parameters can be expressed in terms of these independent parameters such as 
\besub
\bea
M_{\mathcal{F}} &=& m_{1}c_{\theta}^2+m_2s_{\theta}^2 ,   \\
M_{f} &=& m_{1}s_{\theta}^2+m_2c_{\theta}^2, \\
y &=& \frac{\sqrt{2}(m_2-m_{1})c_{\theta}s_{\theta}}{v} = \frac{\sqrt{2}}{v} \Delta_{21} c_{\theta}s_{\theta}.  
\eea
\label{dependent_parameters}
\eesub
\noindent Hence a small $\theta$ would indicate that the lightest eigenstate contains mostly the doublet component. 

\section{Constraints on the Model}
\label{constraints} 
Here we summarize all sorts of constraints (theoretical as well as experimental) to be applicable to the model parameters. 

\subsection{Theoretical constraints}

\begin{itemize}
\item[(i)]{\bf{Stability:}} The scalar potential should be bounded from below in all the field directions of the field space. This leads to the following constraints involving the quartic couplings as 
\besub
\bea
\lambda_H (\mu) ,~ \lambda_S (\mu) \geq  0,  \\
\lambda_{HS}(\mu)~+~\sqrt{\frac{2}{3}\lambda_{H}(\mu)\lambda_{S}(\mu)}~\geq 0, 
\,\, 
\eea
\label{copos} 
\eesub
where $\m$ is the running scale. These conditions should be analysed at all the energy scales  
upto the Planck scale ($M_{\rm Pl}$) in order to maintain the stability of the scalar potential till $M_{\rm Pl}$.

\item[(ii)]{\bf{Perturbativity:}} A perturbative theory demands that the model parameters should obey: 
\bea
|\l_{i}|< 4\pi~ {\rm{and}}~ |g_i|,| y, \a_{1},\a_{2}|<\sqrt{4\pi}  .
\label{pert} 
\eea
where ${g}_i$ and (${y},\a_{1},\a_{2}$) are the SM gauge couplings and Yukawa couplings involving BSM fields respectively. We will ensure the perturbativity of the of the couplings present in the model till the $M_{\rm Pl}$ by employing the renormalisation group equations (RGE). In addition, the perturbative unitarity associated with the S matrix corresponding to $2 \rightarrow 2$ scattering processes involving all two-particle initial and final states \cite{Horejsi:2005da,Bhattacharyya:2015nca} are considered. It turns out that the some of the scalar couplings of
Eq. (\ref{p1}) are bounded by 
\bea
\l_{H}< 4\pi~, \l_{HS}< 8\pi~ {\rm{and}} \nonumber\\
\frac{1}{4} \bigg{(}12\l_{H}+\l_{S}\pm\sqrt{16\l_{HS}^2+(\l_S-12\l_{H})^2})\bigg{)}< 8\pi
\label{pert_unita} 
\eea
\end{itemize}

\subsection{Experimental constraints}
\begin{itemize}
\item[(i)]{\bf{Relic density and Direct detection of DM:}} In order to constrain the parameter space of the model, we use the measured value of the DM relic abundance provided by the Planck experiment\cite{Aghanim:2018eyx} and apply the limits on DM direct detection cross-section from LUX \cite{Akerib:2016vxi}, PandaX-II \cite{Tan:2016zwf,Cui:2017nnn} and XEXON1T \cite{Aprile:2017iyp,Aprile:2018dbl} experiments. Detailed discussion on the dark matter phenomenology involving effects of these constraints is presented in section \ref{DMph}.

\item[(ii)]{\bf{Collider constraints:}}

With the inclusion of additional coloured particles (VLQs) in the present setup, the collider searches can be important. 
In general, the VLQs couple (i) with the SM gauge bosons and (ii) to the SM quarks through Higgs via Yukawa type of interactions if allowed by the symmetries of the construction. Such a mixing, after electroweak symmetry breaking, between the VLQs (generically denoted by $T$ say) and SM quarks  (say $q$) induces a coupling involving $TqX$, with $X \equiv$ gauge bosons 
or higgs. This new interaction and mixing with the SM quarks open up various production as well as the decay modes of these VLQs at LHC which in turn provide stringent constraints on their masses and their couplings. In fact, from the $pp$ collision at LHC, VLQs can be pair produced at LHC by strong, weak and electromagnetic interaction (as appropriate). It can also be singly produced via interactions $TqX$, in association with gauge boson or SM quark in the final state. The dominant production 
channel depends crucially upon with which generation(s) of SM quarks, the VLQs are allowed to couple. 

Once produced, each VLQ decays to a SM quark and a gauge boson or Higgs. Such a decay resulting a large number of jets 
only, can not be traced back efficiently as the VLQ-origin due to immense QCD background already present in $pp$ collision. However, note that in case $W$ bosons are expected to be present in the final state either from the decay of VLQ or from a subsequent decay of a $t$ quark, VLQ decays lead to the most interesting signature at LHC with transverse missing energy ($\slashed E_T$), one lepton and jets (VLQs are pair produced). A recent detailed study in this direction \cite{Buckley:2020wzk} 
shows that VLQ mass below $\sim$ 1.5 TeV is almost ruled out with 13 TeV LHC data.

However, the above constraint on the mass of the VLQ becomes less stringent in case the VLQ does not have a direct 
Yukawa coupling with the SM quarks via Higgs as in point (ii) above. Such a situation can be obtained once the VLQ is charged 
under some discrete symmetry (for example odd under an unbroken $Z_2$ as in our case) while SM particles are not. 
Obviously this significantly reduces the number of production and decay channels of the VLQ due to the absence of 
VLQ-SM quark mixing. Usually such a construction is related to the presence of a scalar singlet DM (say $S$ as in 
our case), non-trivially charged under the same discrete symmetry as in \cite{Giacchino:2015hvk,Baek:2016lnv,Baek:2017ykw,Colucci:2018vxz}. In that case, a new portal coupling involving 
VLQ, scalar singlet and the SM quarks is expected though. Then the decay of VLQs would proceed through the missing 
energy signal along with the jets provided enough phase space is available. A recent study of this kind of model 
(top-phillic, $i.e.$ VLQ being singlet couples with S and $t_R$) with LHC run-2 data \cite{Colucci:2018vxz} suggests that the VLQ 
masses below 1 TeV are excluded only if the DM mass is light enough compared to the VLQ to allow such a decay to happen, $e.g$ $m_{VLQ} - m_S \gg m_t$. On the contrary, if the mass difference falls below $m_t$, the constraint becomes much 
weaker and it is shown to allow $m_S$ above 350 GeV with VLQ masses above $\sim$ 400 GeV \cite{Colucci:2018vxz} or so. 

The other possibility where a VLQ couples to lighter SM quark through the scalar singlet DM is extensively 
investigated in \cite{Giacchino:2015hvk} where, 
it is shown that the jets arising from the decay of the VLQs are too soft to be detected at ATLAS or CMS if the VLQs 
couple to the lighter SM quarks. In that case, inclusion of one or more hard jets (due to radiation of gluons from initial/final/intermediate state) is necessary for the detection of the events. Hence multi-jets + $\slashed E_T$ can provide a 
more practical constraints than those obtained from the lowest order two-jet+ $\slashed E_T$ signals. Applying the combined results of multi-jets + $\slashed E_T$ at ATLAS searches together with the constraints coming from the DM direct detection as well as the indirect detection experiments using $\g-$rays, the study in \cite{Giacchino:2015hvk} allows real scalar singlet 
DM mass above 300 GeV with $m_{VLQ}/m_S \simeq 1.2$ indicative of VLQ mass $\sim$ 400 GeV. Also for $m_S$ = 
500 GeV, VLQ having mass 550-750 GeV turns out to be in the allowed range. 

Note that the above correlation between mass of the VLQ and $m_S$ is primarily due to the presence of large Yukawa coupling between VLQs, DM and the SM quark which is also required to satisfy the DM relic density in the work of \cite{Giacchino:2015hvk}. Instead in our case, it turns out that this particular coupling plays only subdominant role in dark matter phenomenology. We have the freedom of choosing smaller Yukawa coupling (compared to \cite{Giacchino:2015hvk}) as in our case the correct relic is effectively produced by the annihilation of VLQs (note that we have both the singlet and doublet VLQs) and the Yukawa coupling $\alpha$ can be restricted to be small. Hence limits obtained on $m_1/m_S$ in \cite{Giacchino:2015hvk} does not apply here. On the other hand, we work with relatively lighter VLQ, we maintain the mass-splitting between the lightest VLQ and the DM to be smaller than $m_t$ so that constraints from the VLQ decay associated with jets+$\slashed E_T$+charged lepton would not be applicable here. Even then, we use the VLQ's mass around 500 GeV as a conservative consideration for their mass. A weaker bound on VLQ's mass however prevails as $m_{\rm VLQ} > 100$ GeV similar to the constraints in case of squark searches, from LEP experiment \cite{Abbiendi:2002mp}.  
\end{itemize}

\section{Dark Matter Phenomenology}
\label{DMph}

In this section, we elaborate on the strategy to calculate the relic density and the direct detection of the dark matter $S$ in the present setup. 
As we have discussed, apart from the scalar singlet $S$ the present setup 
also incorporates additional fermions, in the form of one vector like quark doublet $\mathcal{F}$ and one vector like quark singlet $f'$. These VLQs talk to the SM via its ($i$) Yukawa interactions with (a) the dark matter $S$, 
(b) the SM Higgs and ($ii$) gauge interaction ($e.g.,$ with gluons). 
Below we provide a list of relevant Yukawa interaction vertices (belonging to ($i$) as above). 
\bea
\nonumber
S \overline{\mathcal{F}_1} u &:& \frac{1}{2}[\a_1 c_{\theta}(1-\g_5)-\a_2 s_{\theta}(1+\g_5)], \\
 \nonumber
S \overline{\mathcal{F}_2} d &:& \frac{1}{2}[\a_1(1-\g_5)], \\
S \overline{f} u &:& \frac{1}{2}[\a_1 s_{\theta}(1-\g_5)+\a_2 c_{\theta}(1+\g_5)], ~~{\rm{and}}\\
\nonumber
h \overline{\mathcal{F}_1} f &:& \frac{y}{\sqrt{2}}(1-2s_{\theta}^2), \\
\nonumber
h \overline{\mathcal{F}_1} \mathcal{F}_1 &:& -\frac{y}{\sqrt{2}}c_{\theta}s_{\theta}, \\
h \overline{f} f &:& \frac{y}{\sqrt{2}}c_{\theta}s_{\theta},
\eea
where $u$ represents all the up type SM quarks and $d$ represents all the down type SM quarks.

\begin{figure}[]
\centering
\subfigure[]{
\includegraphics[scale=0.40]{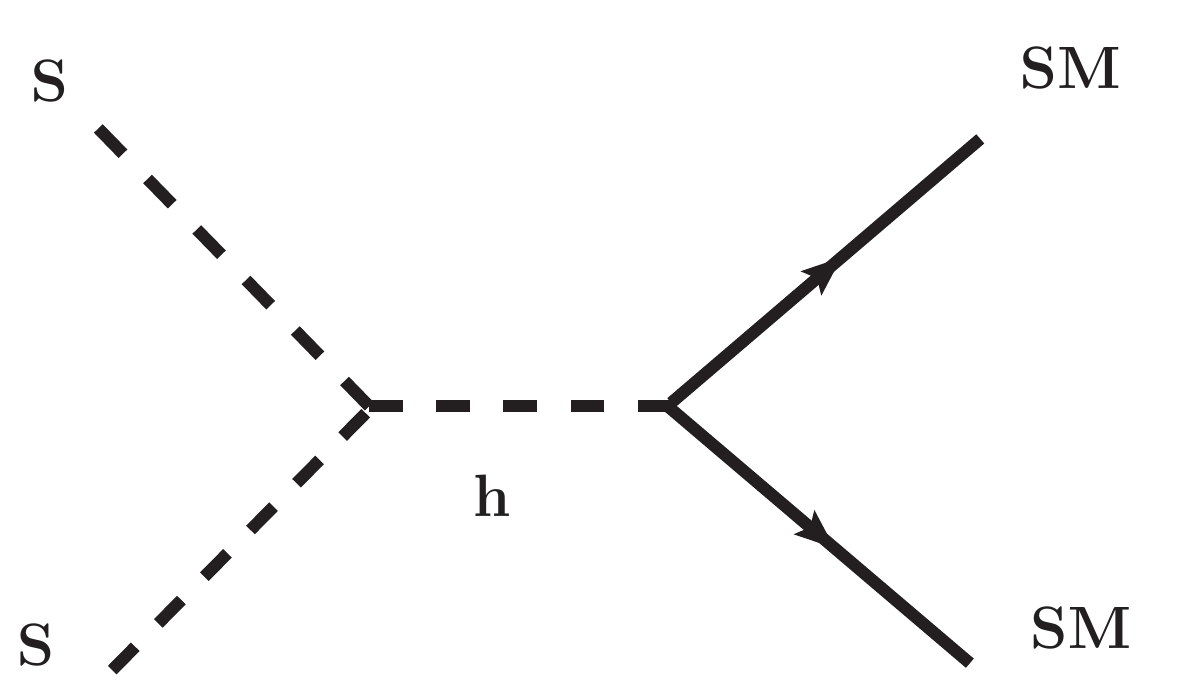}}
\subfigure[]{
\includegraphics[scale=0.40]{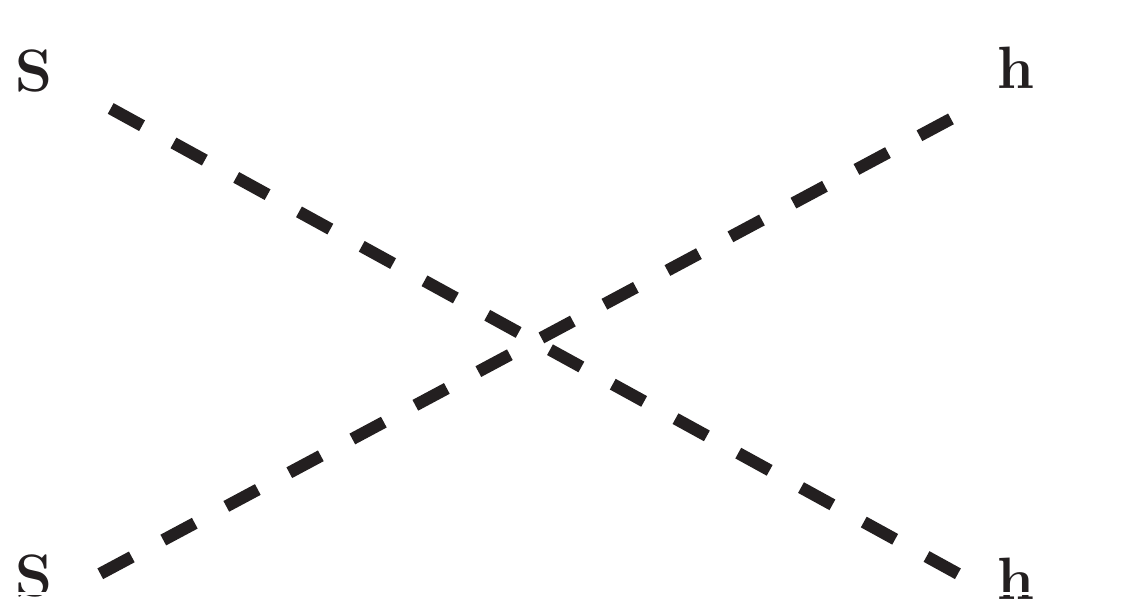}}
\subfigure[]{
\includegraphics[scale=0.40]{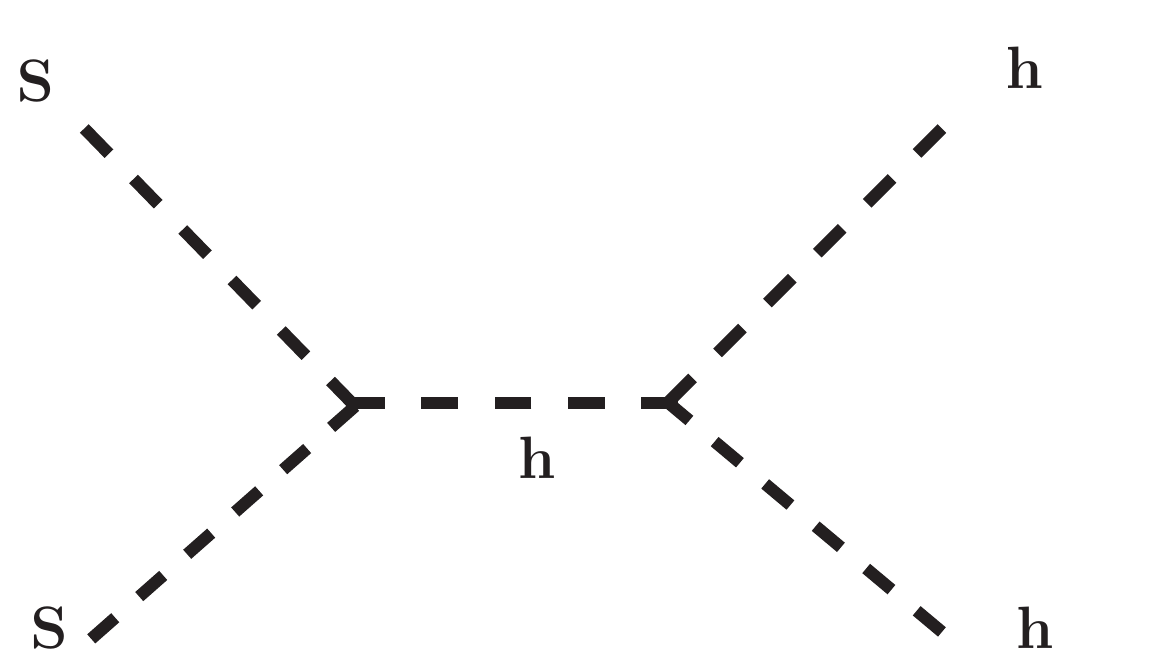}}
\subfigure[]{
\includegraphics[scale=0.40]{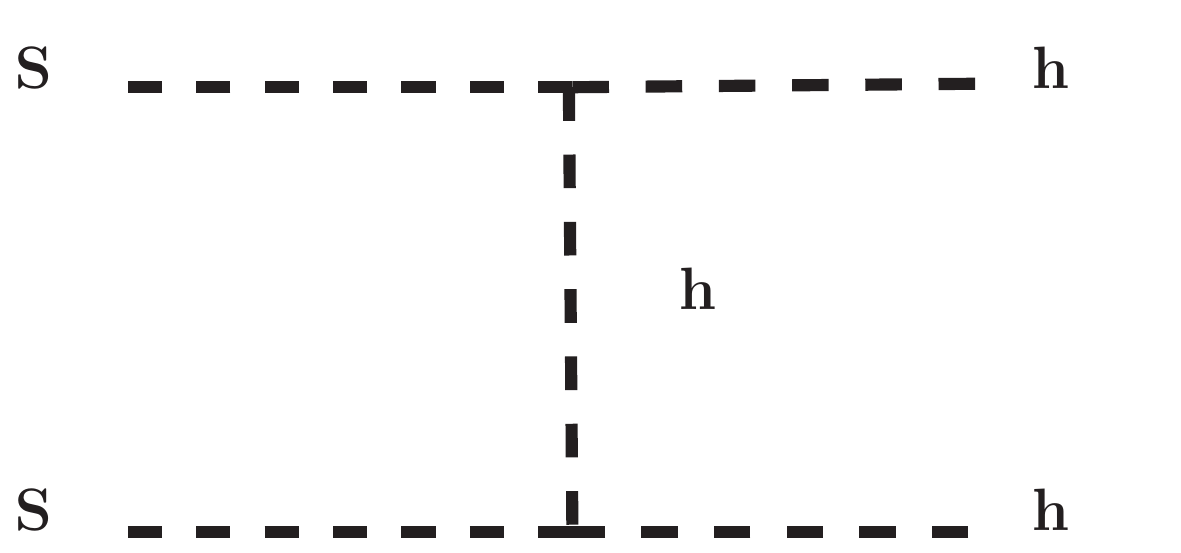}}
\subfigure[]{
\includegraphics[scale=0.40]{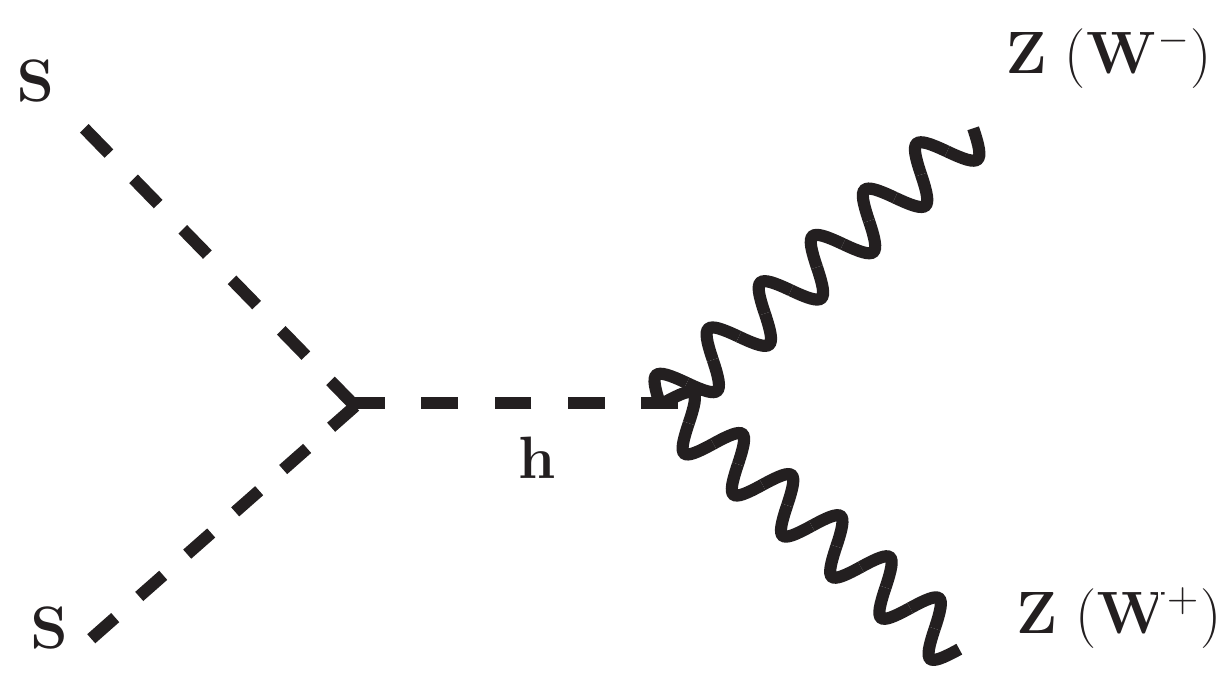}}
\subfigure[]{
\includegraphics[scale=0.40]{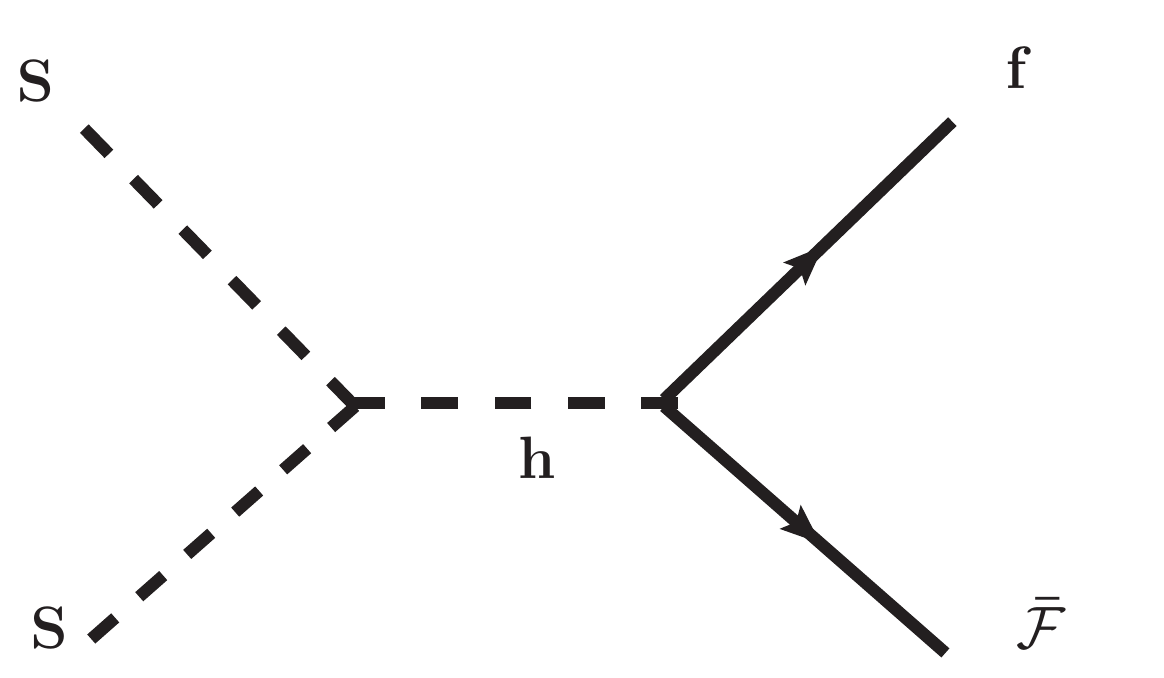}}
\subfigure[]{
\includegraphics[scale=0.40]{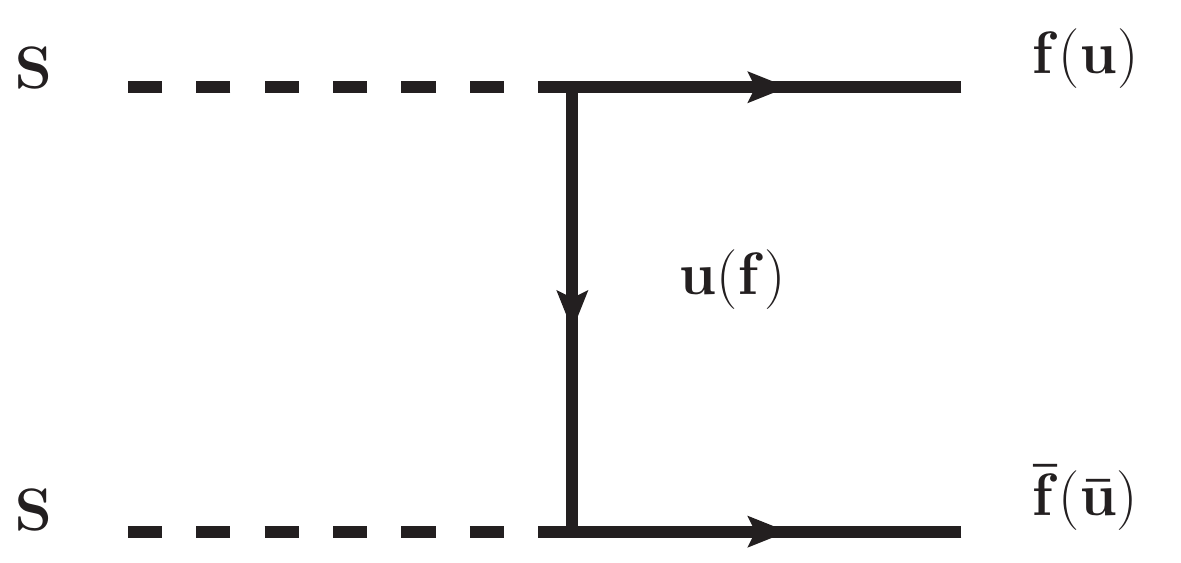}}
\subfigure[]{
\includegraphics[scale=0.40]{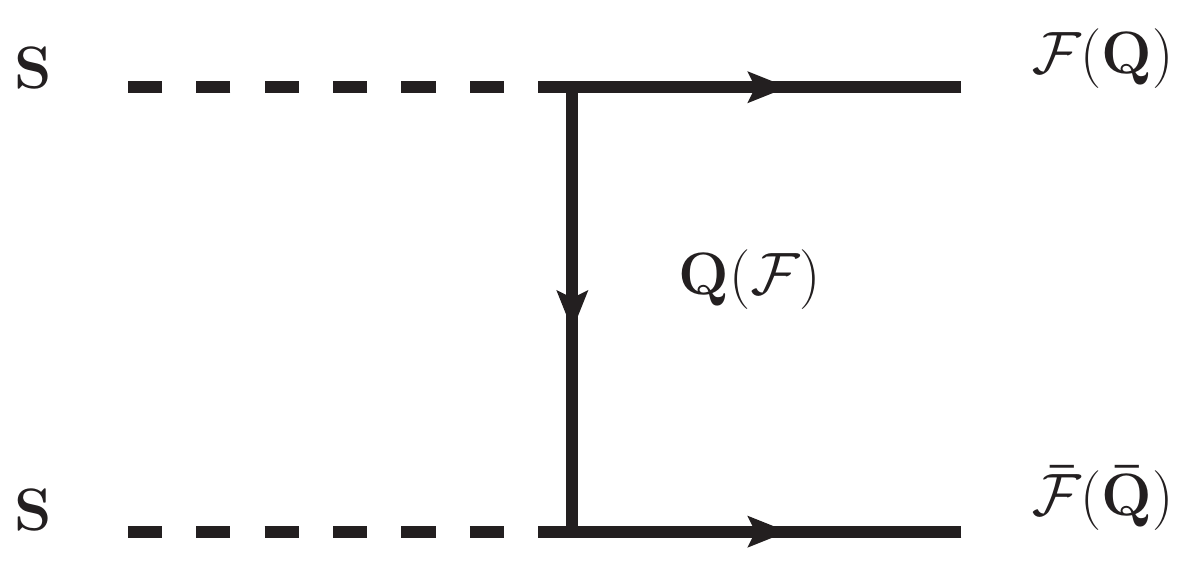}}
\caption{Annihilation channels for scalar singlet dark matter $S$.}
\label{feynS}
\end{figure}
\begin{figure}[]
\centering
\subfigure[]{
\includegraphics[scale=0.40]{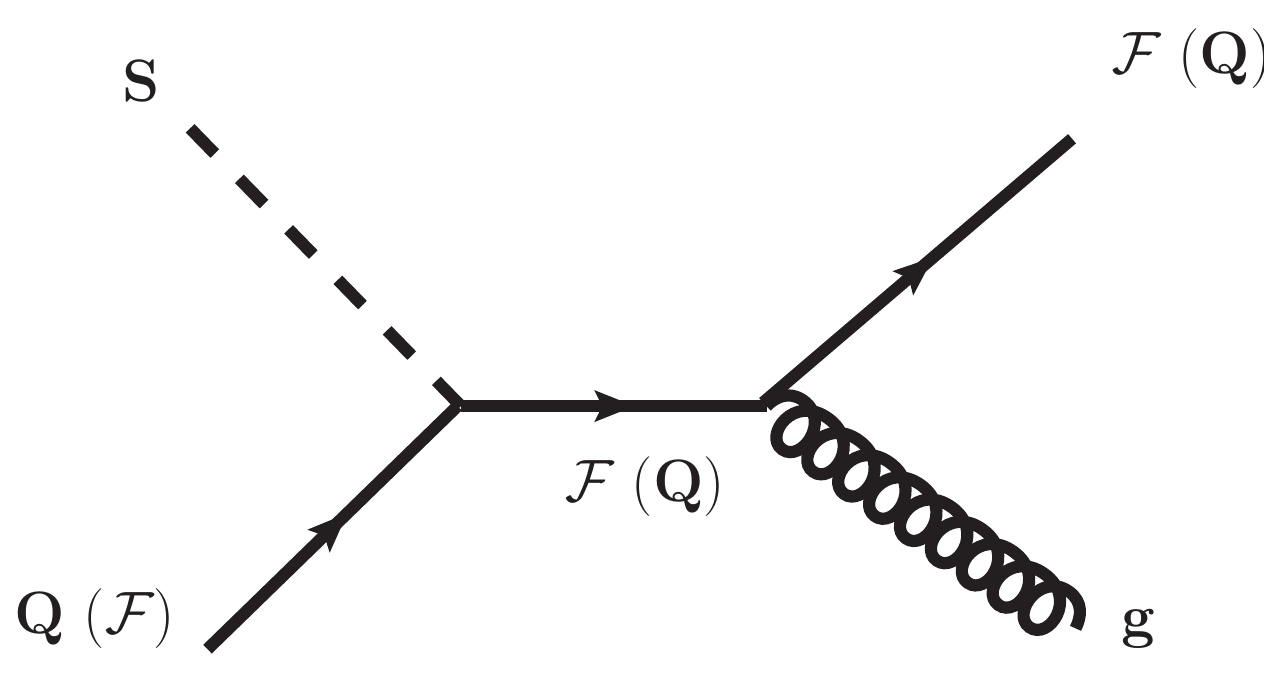}}
\subfigure[]{
\includegraphics[scale=0.40]{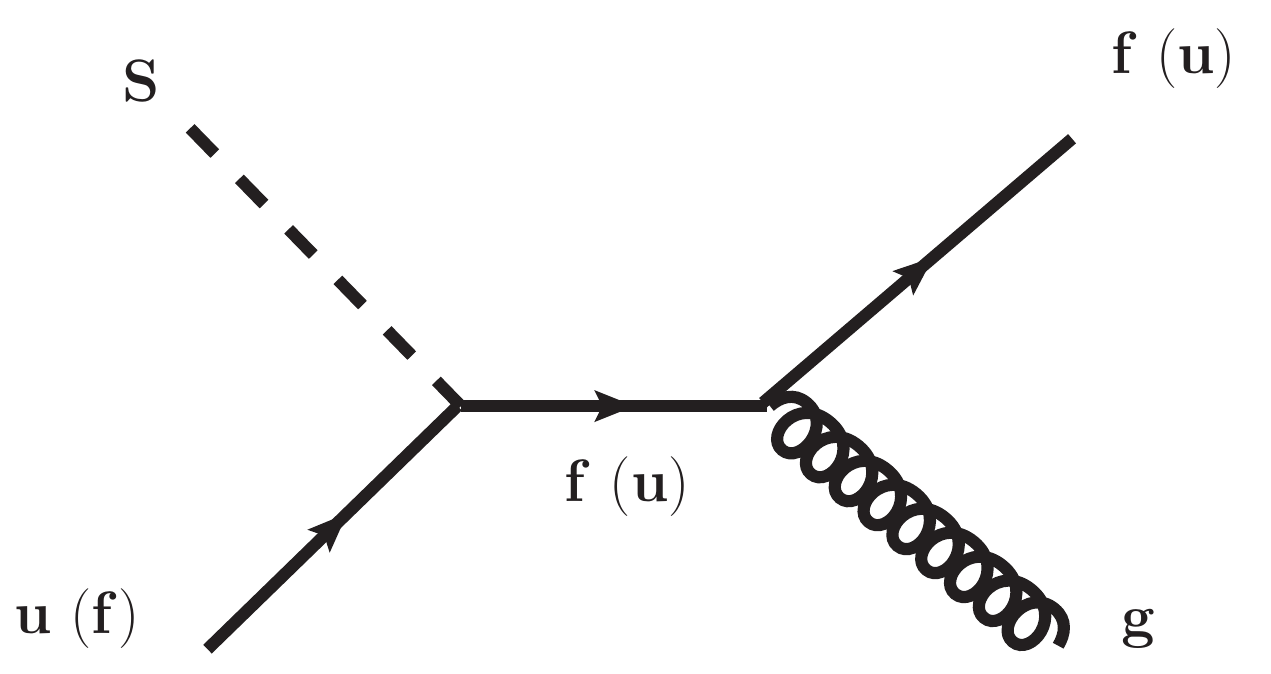}}
\subfigure[]{
\includegraphics[scale=0.40]{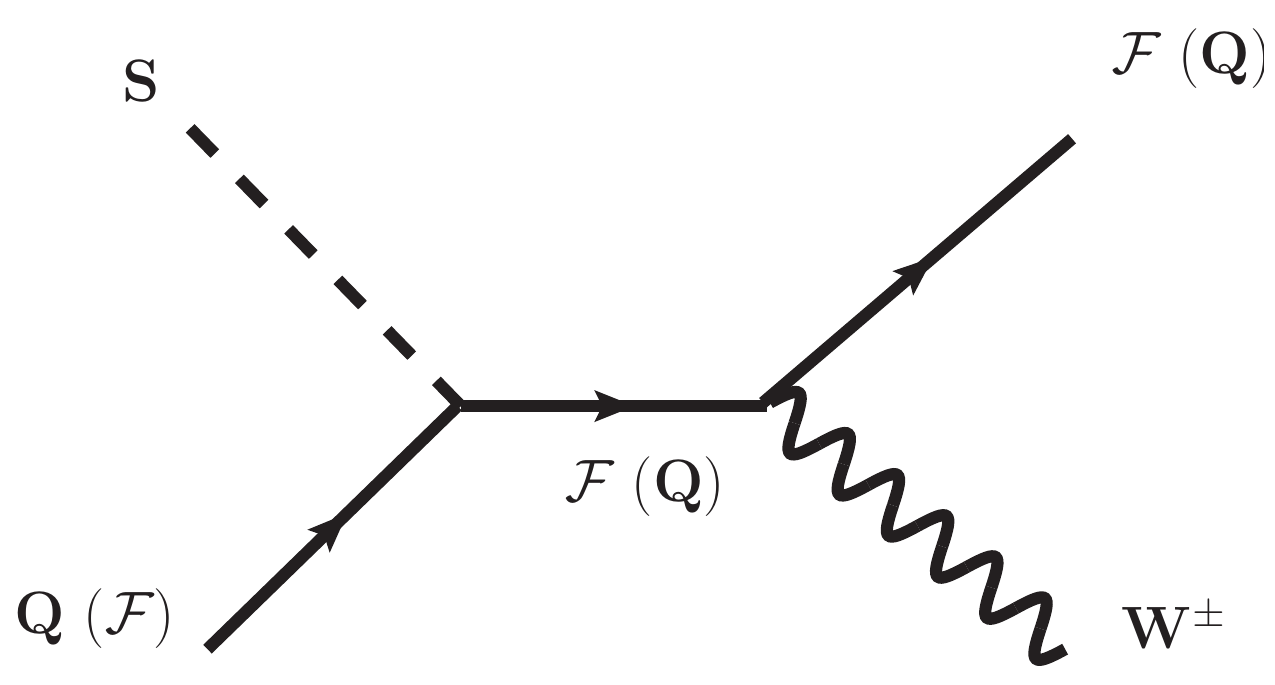}}
\subfigure[]{
\includegraphics[scale=0.40]{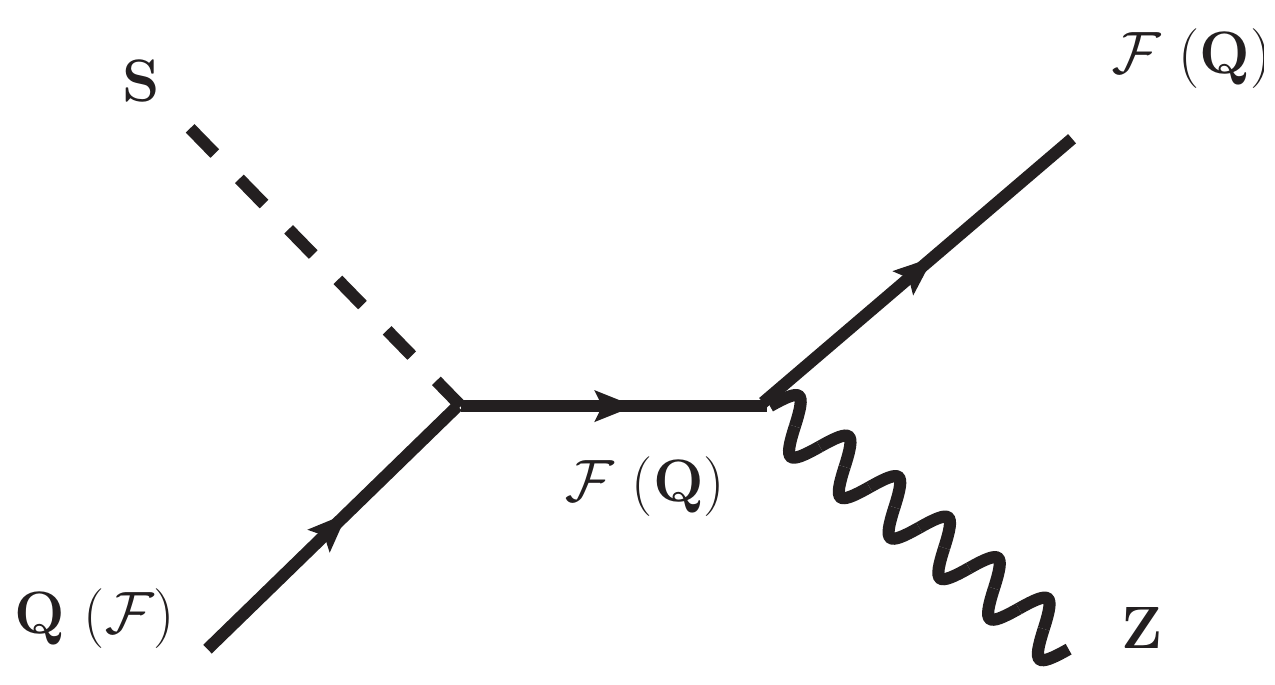}}
\subfigure[]{
\includegraphics[scale=0.40]{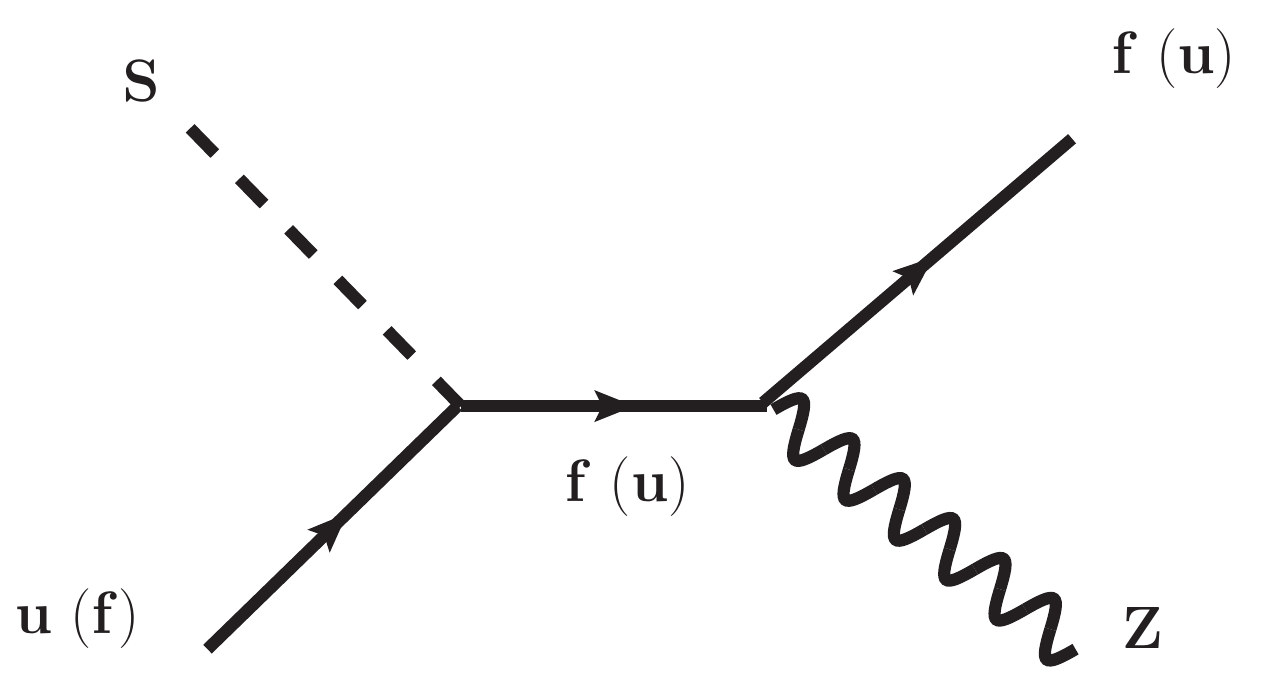}}
\subfigure[]{
\includegraphics[scale=0.40]{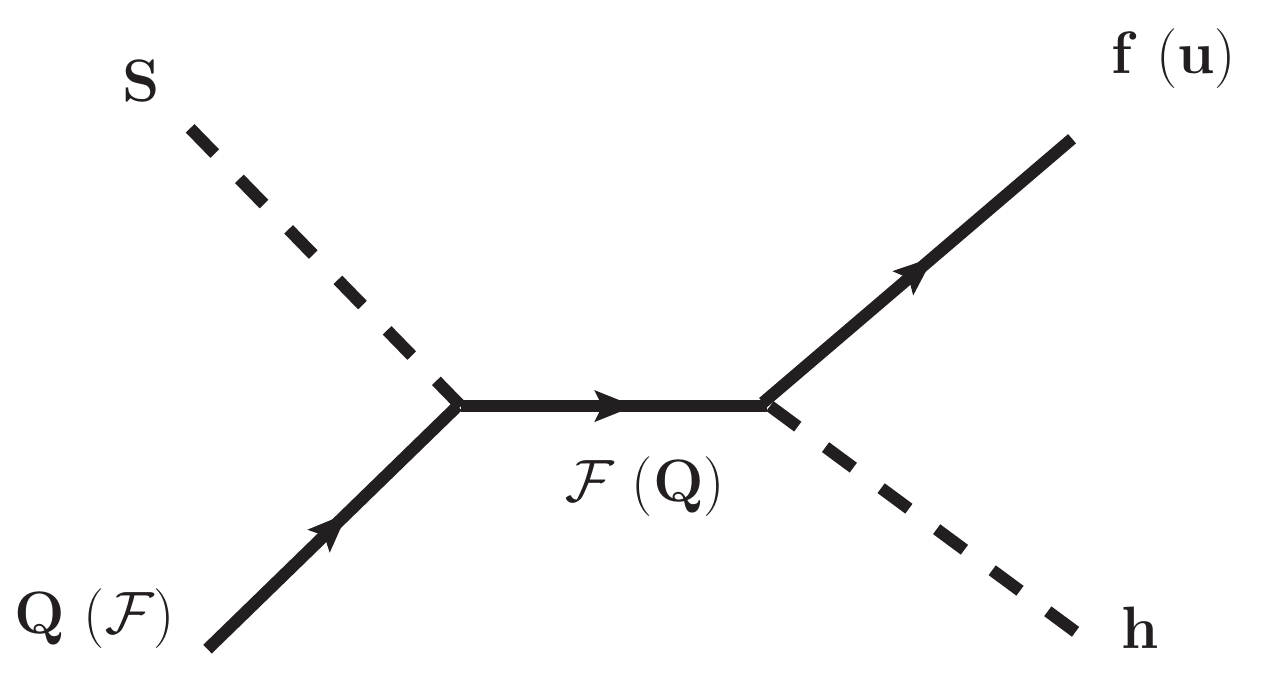}}
\subfigure[]{
\includegraphics[scale=0.40]{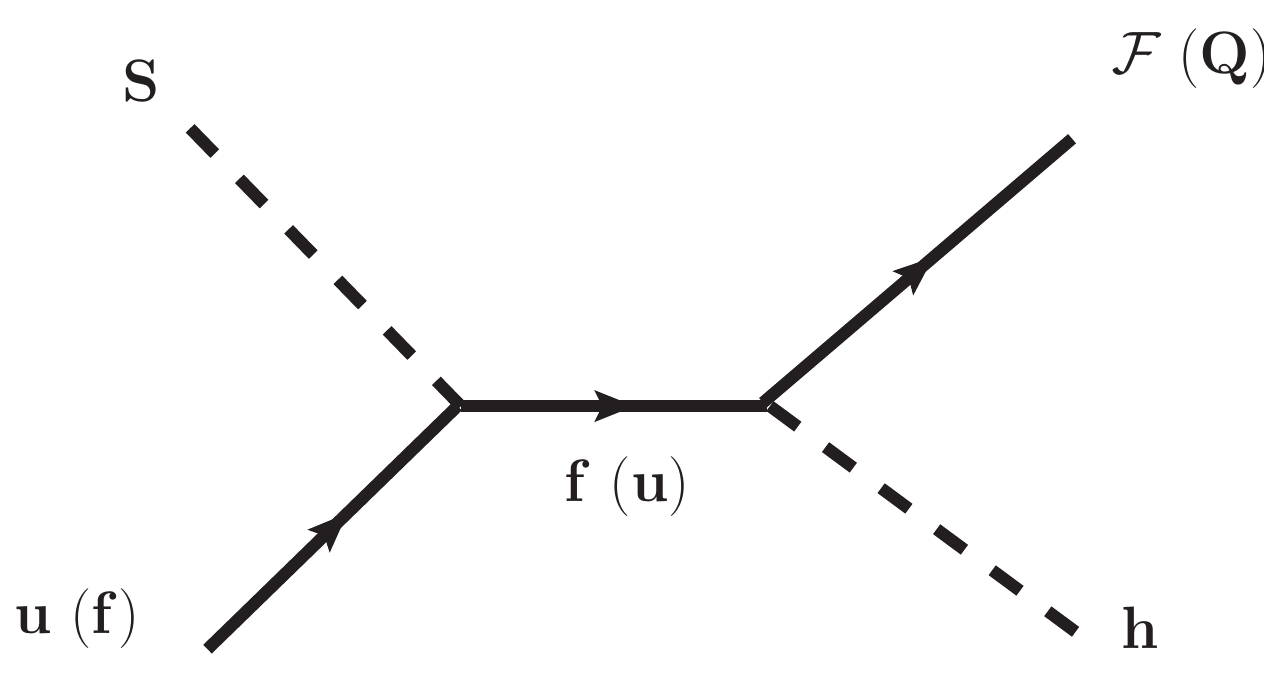}}
\caption{Coannihilation channels for scalar singlet dark matter $S$.}
\label{coann}
\end{figure}
\begin{figure}[]
\centering
\subfigure[]{
\includegraphics[scale=0.40]{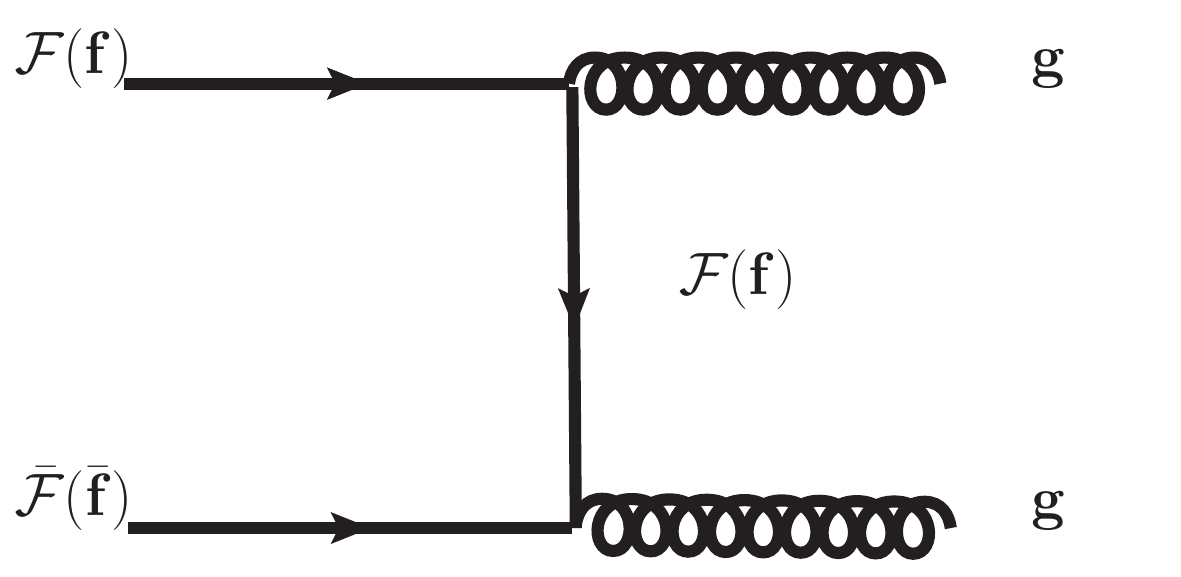}}
\subfigure[]{
\includegraphics[scale=0.40]{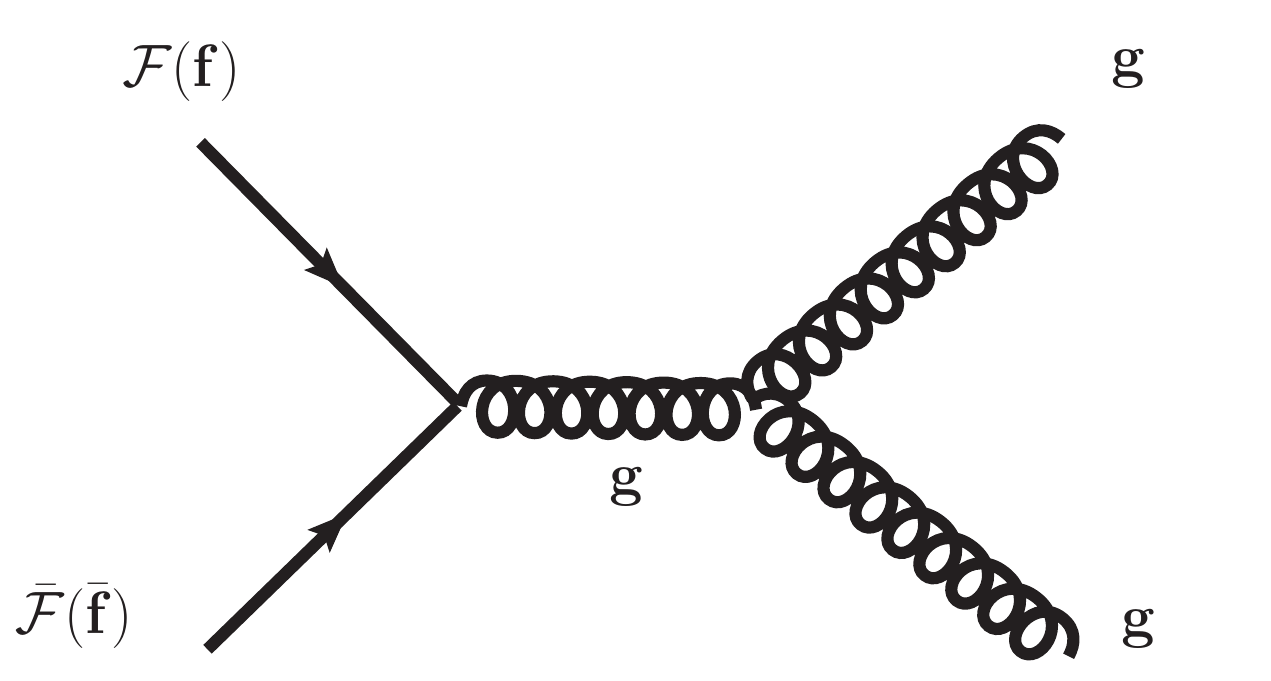}}
\subfigure[]{
\includegraphics[scale=0.40]{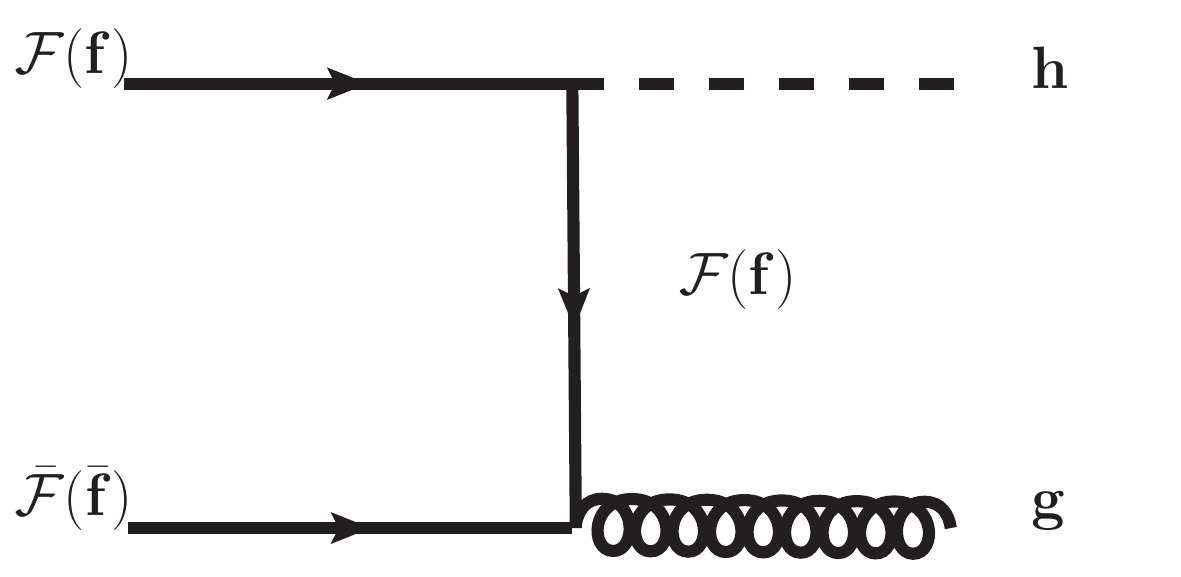}}
\subfigure[]{
\includegraphics[scale=0.40]{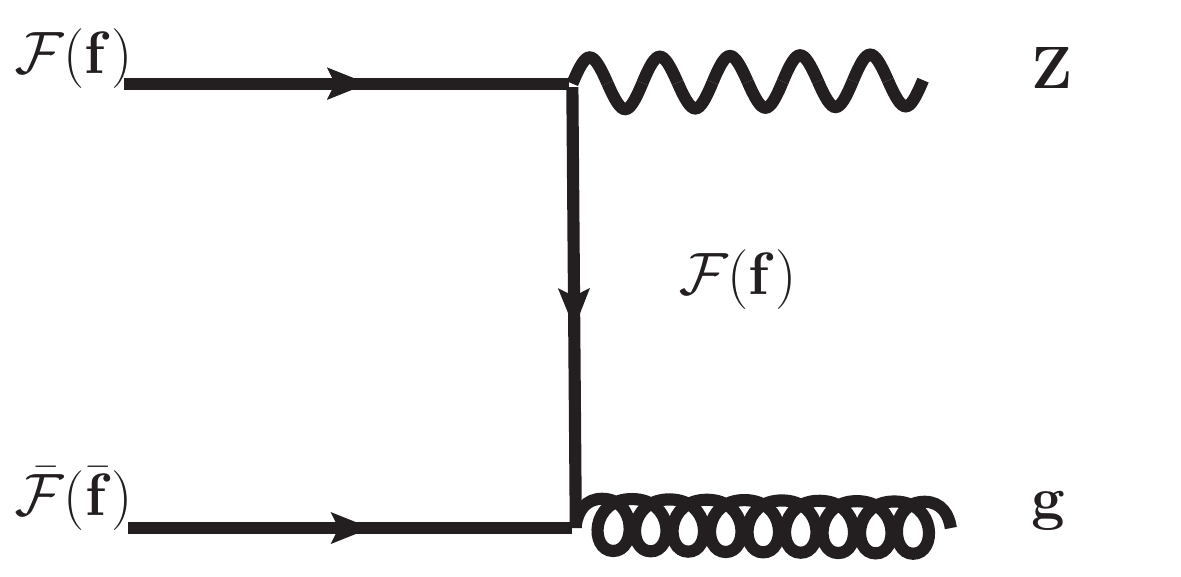}}
\subfigure[]{
\includegraphics[scale=0.40]{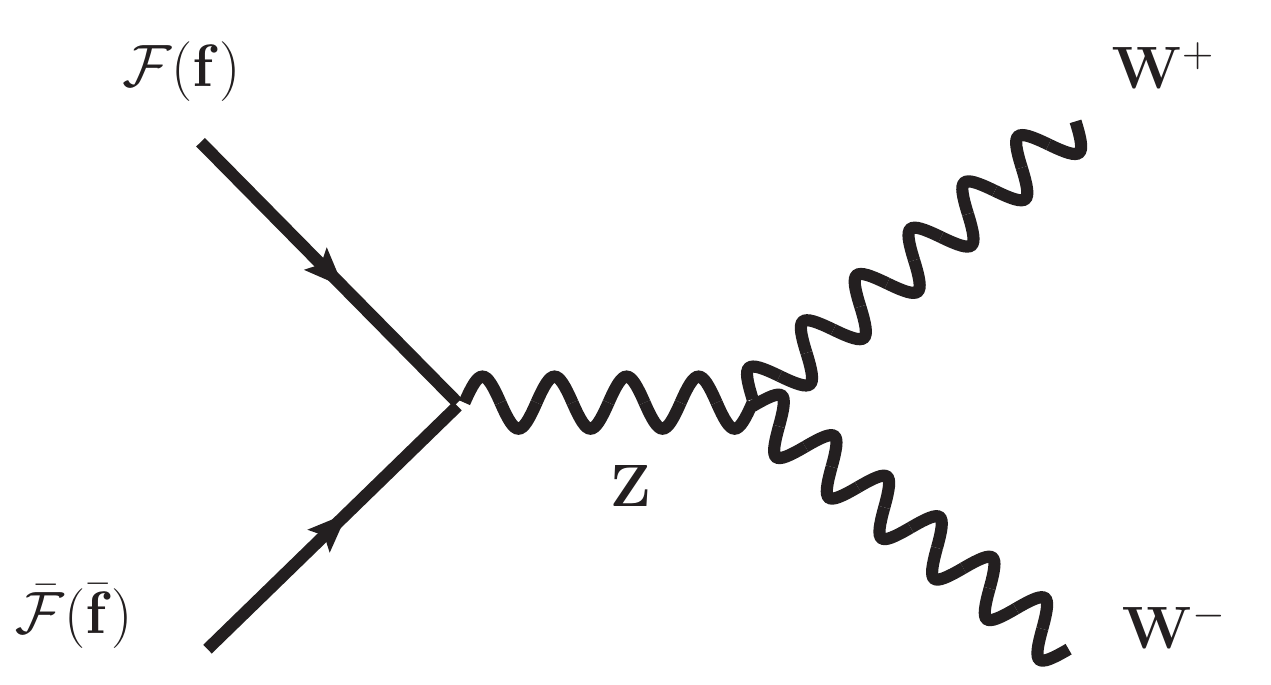}}
\subfigure[]{
\includegraphics[scale=0.40]{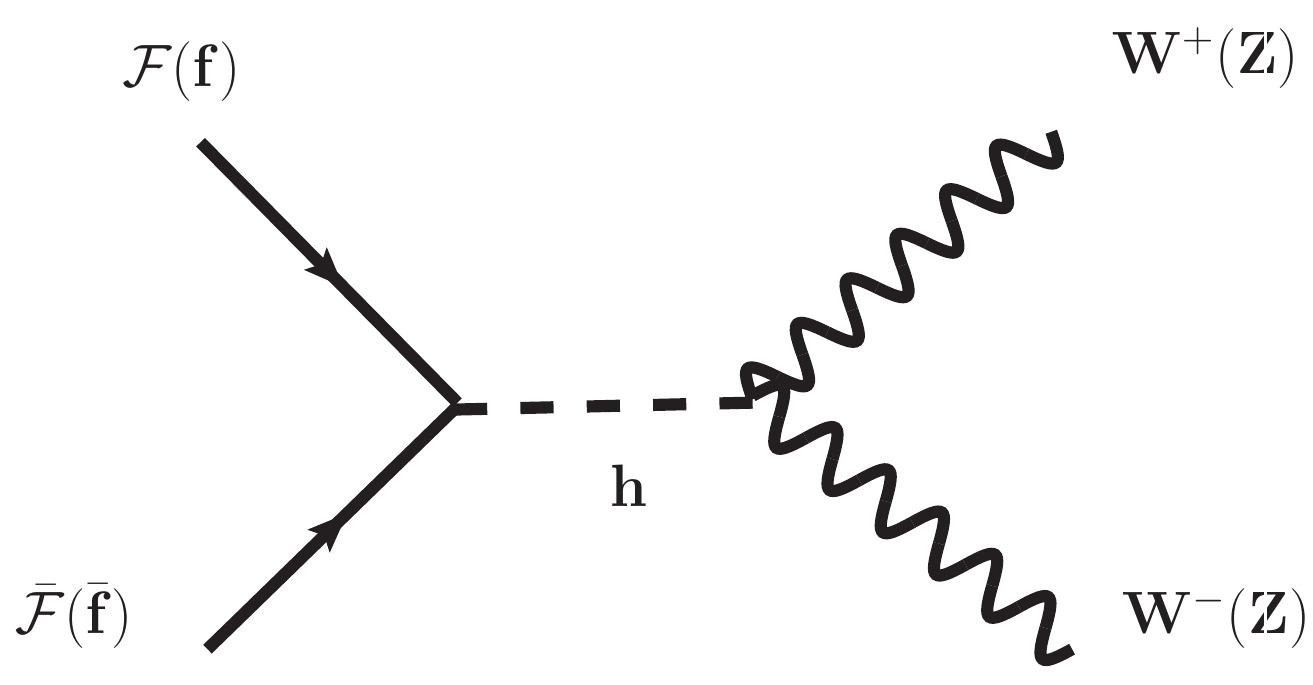}}
\subfigure[]{
\includegraphics[scale=0.40]{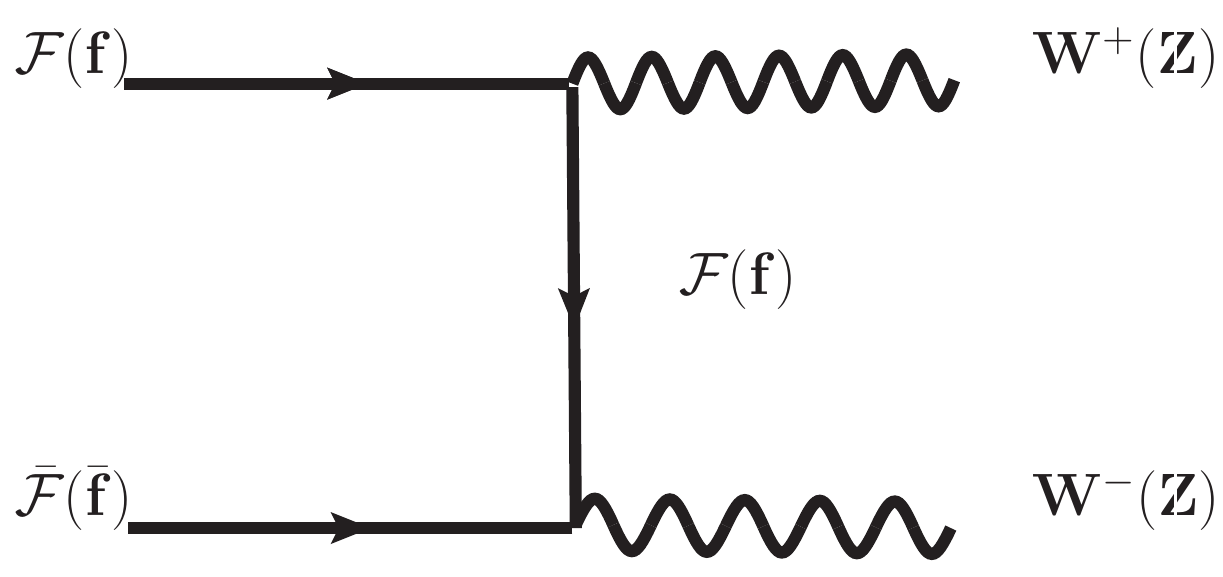}}
\subfigure[]{
\includegraphics[scale=0.40]{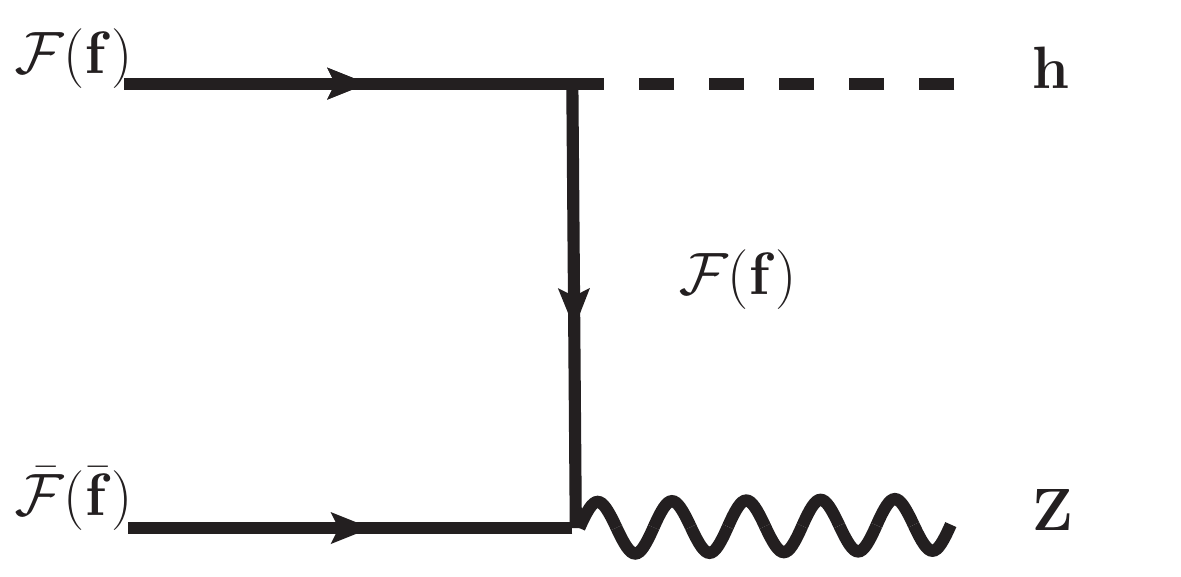}}
\subfigure[]{
\includegraphics[scale=0.40]{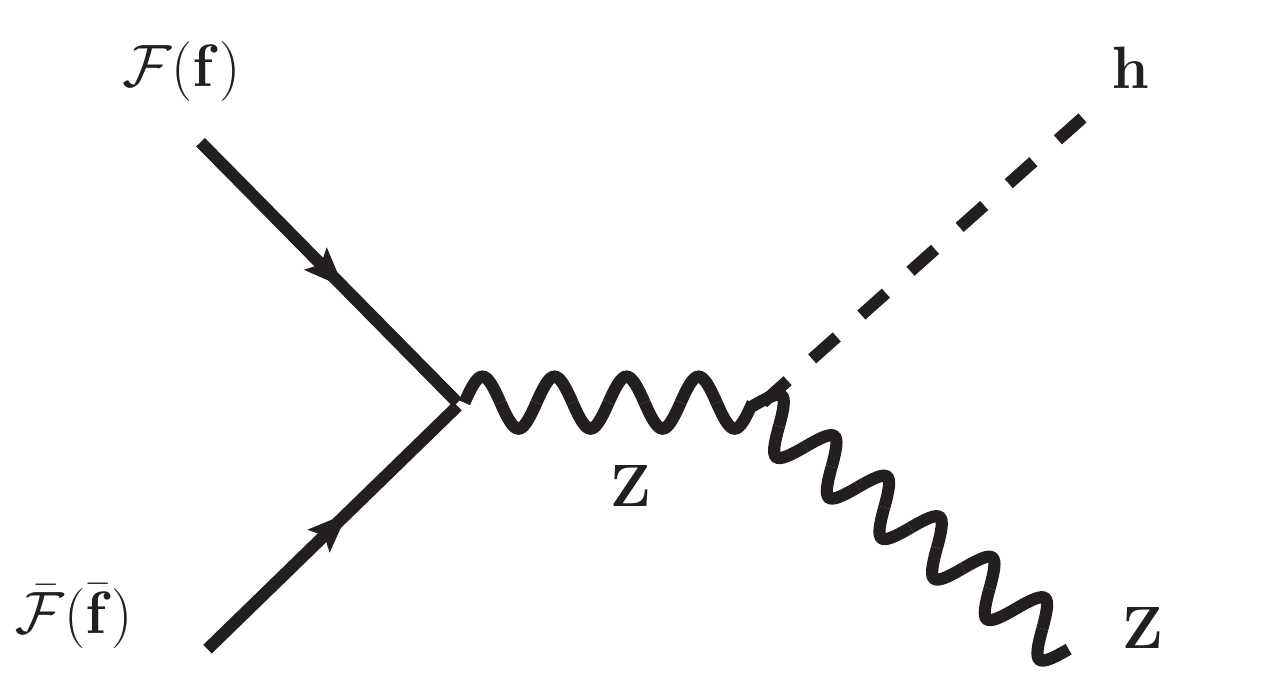}}
\subfigure[]{
\includegraphics[scale=0.40]{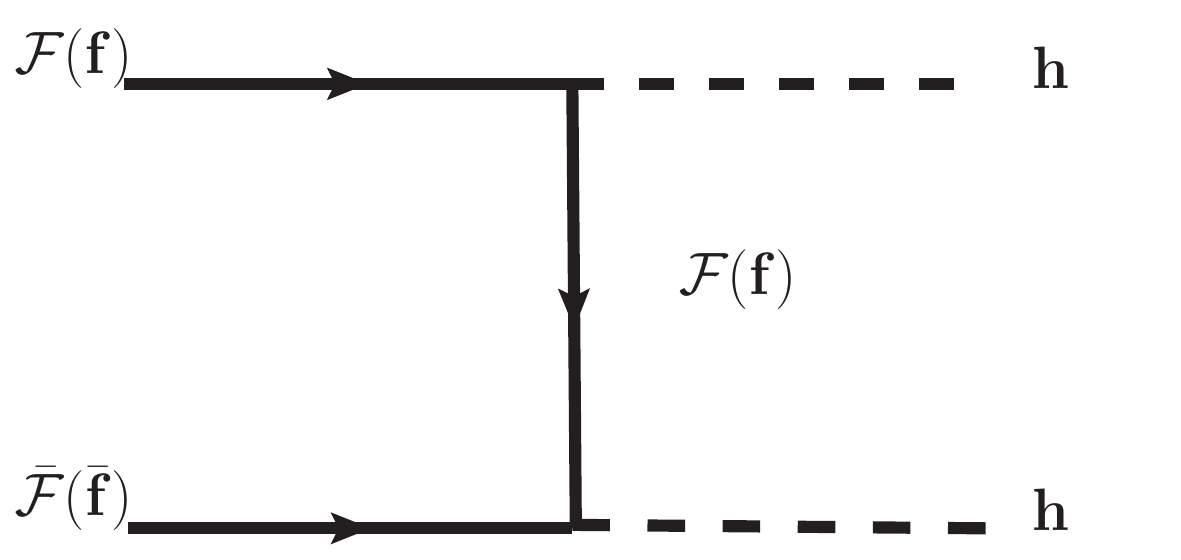}}
\subfigure[]{
\includegraphics[scale=0.40]{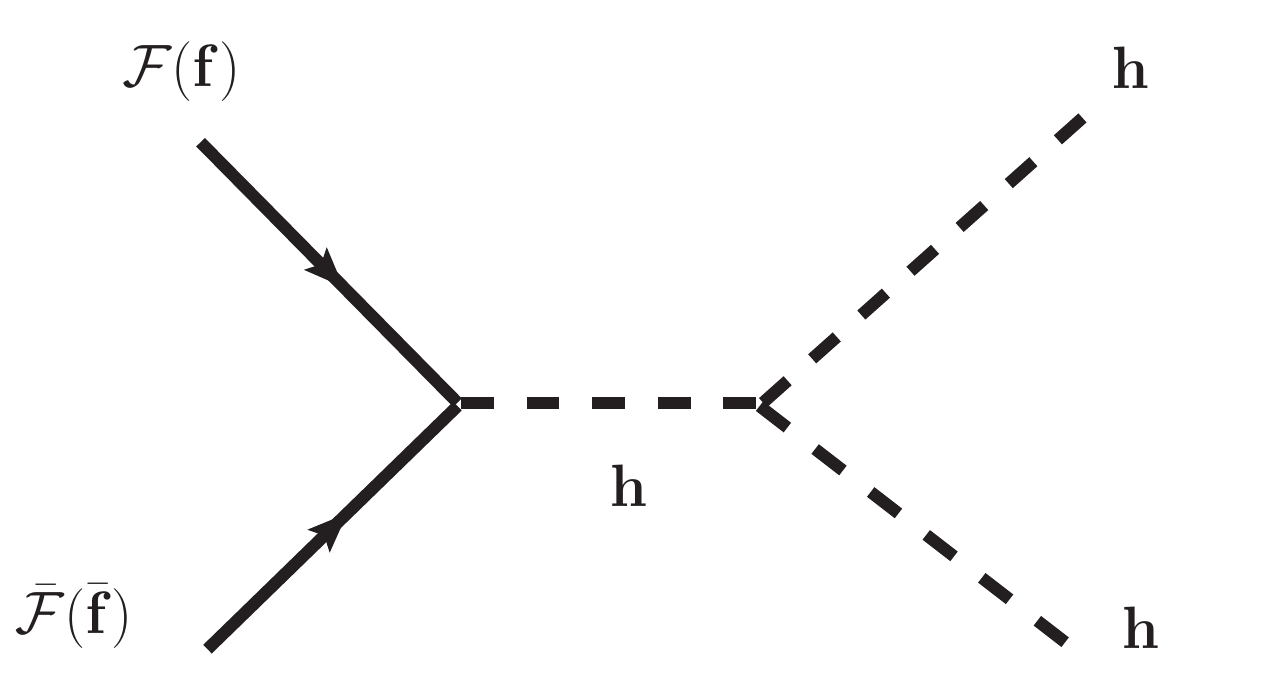}}
\subfigure[]{
\includegraphics[scale=0.40]{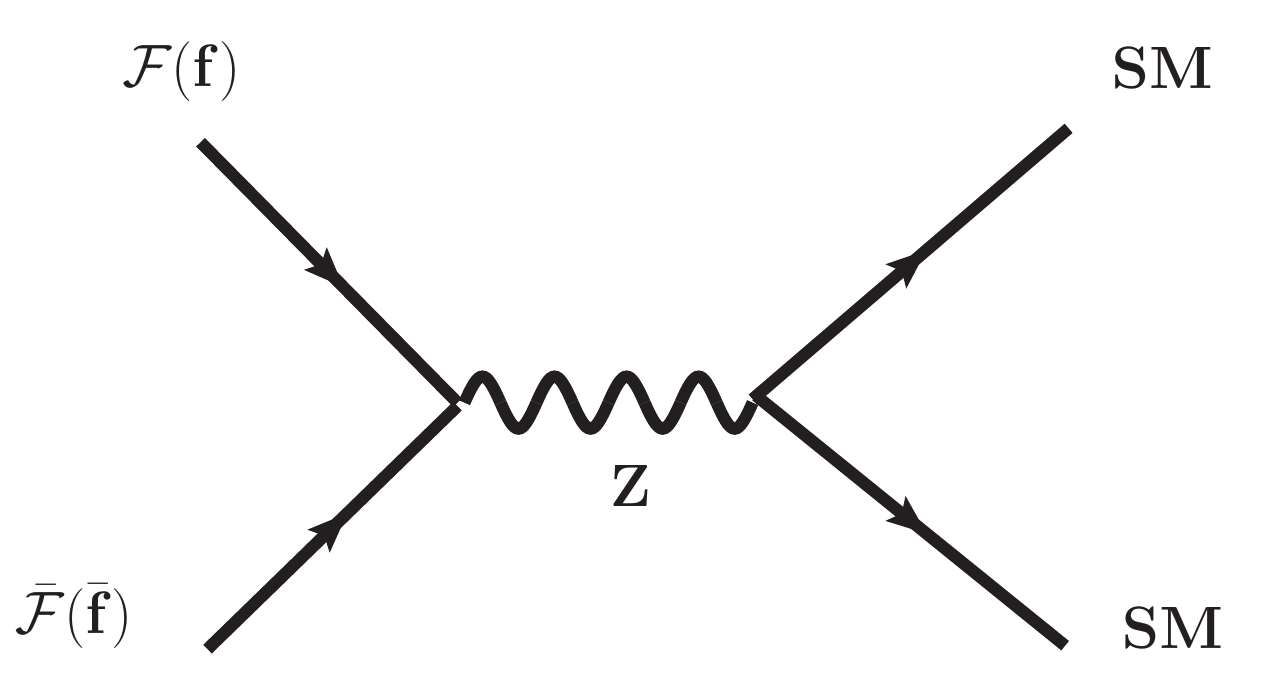}}
\subfigure[]{
\includegraphics[scale=0.40]{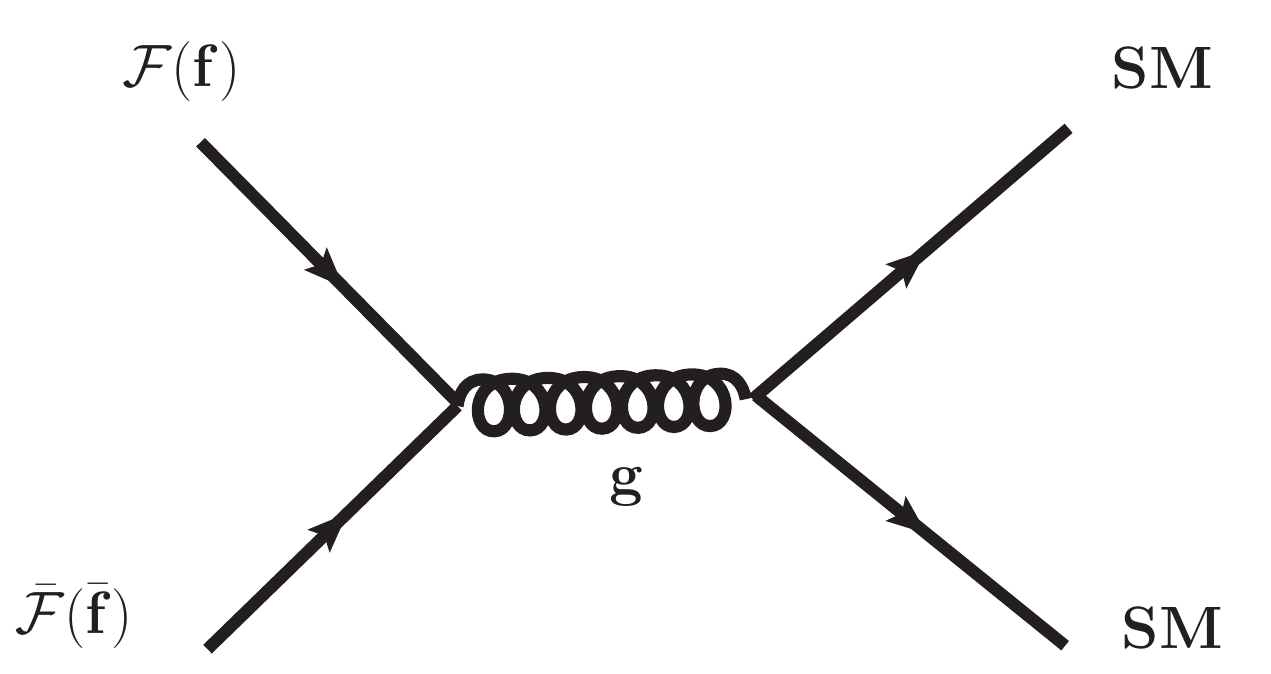}}
\subfigure[]{
\includegraphics[scale=0.40]{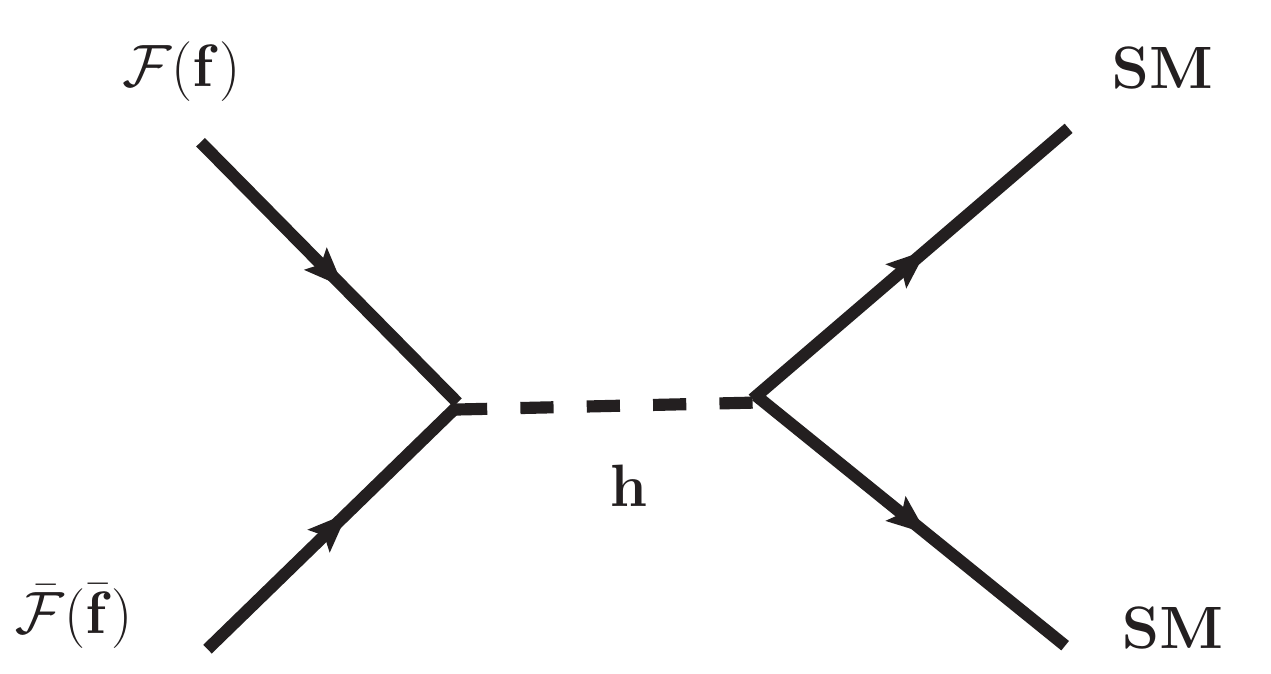}}
\subfigure[]{
\includegraphics[scale=0.40]{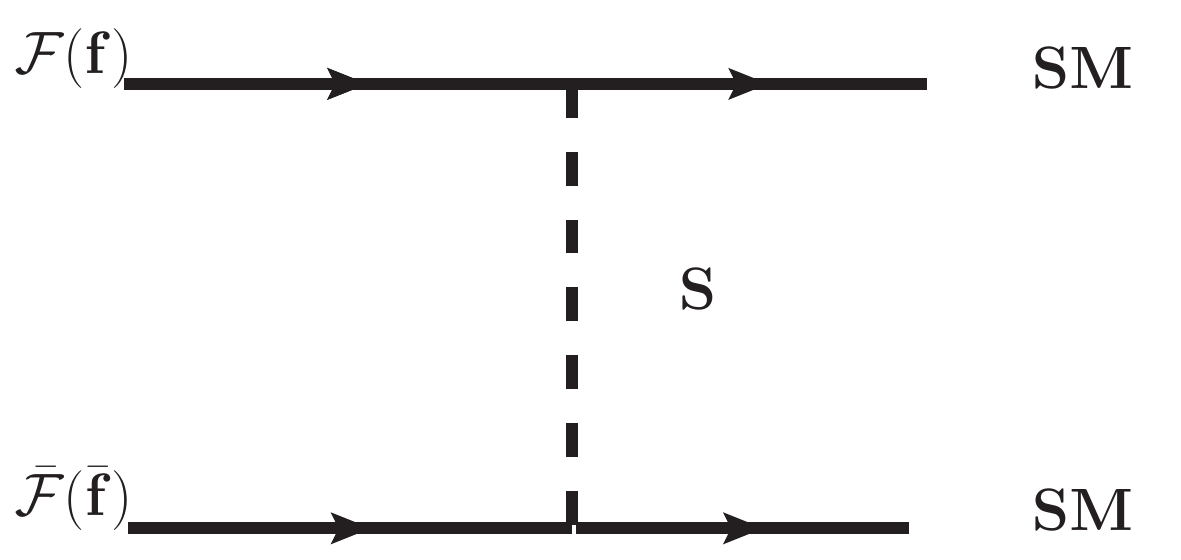}}
\subfigure[]{
\includegraphics[scale=0.40]{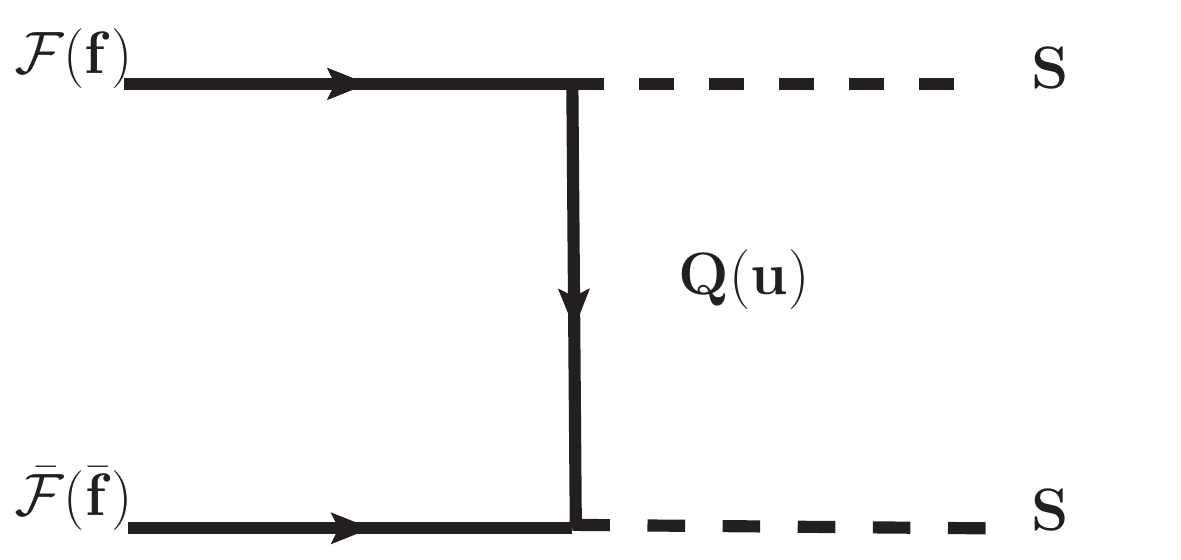}}
\subfigure[]{
\includegraphics[scale=0.40]{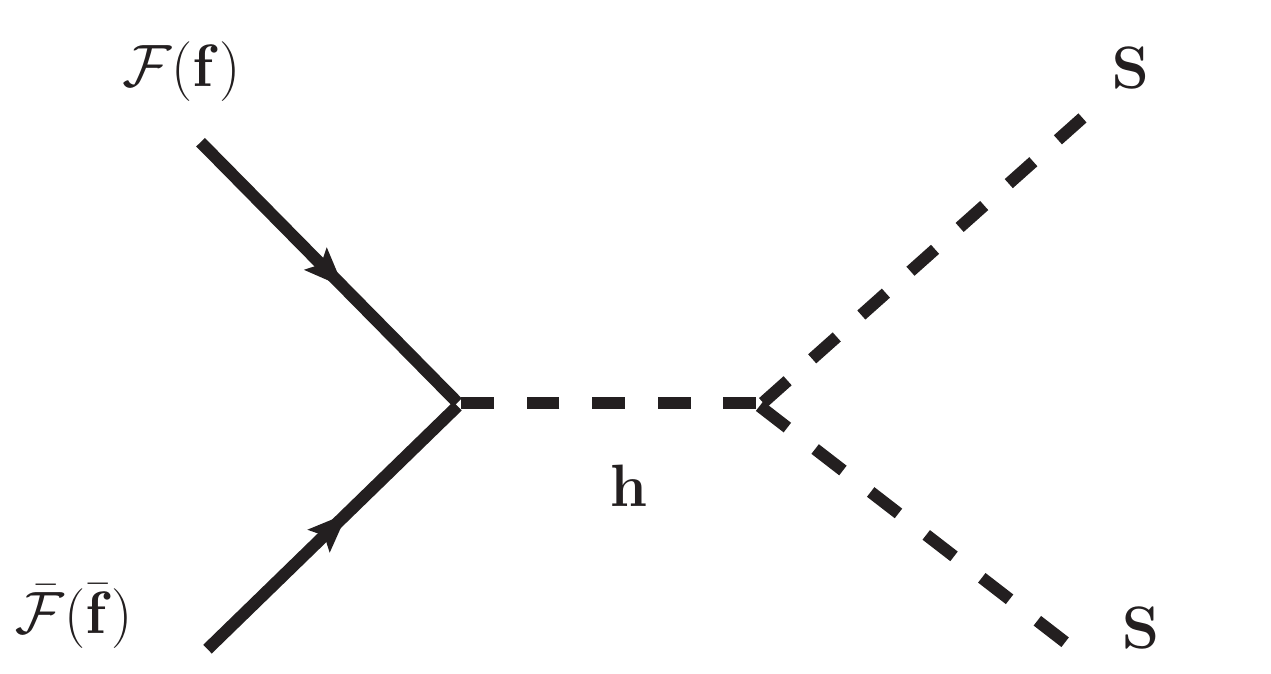}}
\caption{Annihilation channels for vector like quarks.}
\label{feynF}
\end{figure}

Presence of these additional interactions opens up new possiblities for the singlet scalar dark matter to 
co-annihilate with the new VLQs into the SM final states. Now, in order to study the evolution of the DM in 
the universe one needs to solve the Boltzmann equation of the DM. Before going into the details of the 
Boltzmann equation, we first identify and categorise different annihilation and co-annihilation channels of 
the dark matter $S$. In Fig. \ref{feynS} we show all the possible annihilation channels of the dark matter 
whereas in Fig. \ref{coann} we show only the additional co-annihilation channels  which come into the picture 
due to the presence of additional fermions discussed above and finally in Fig. \ref{feynF} we show all the possible annihilation channels of the VLQs in the present setup. 
Considering the DM $S$ as the lightest field among the $Z_2$ odd ones, diagrams involving VLQs in the final state will not contribute in the annihilation process. 

\subsection{Methodology}

As discussed above, the co-annihilation of the dark matter $S$ with the VLQs would play a crucial role 
in explaining the relic density. It has been also shown in \cite{Giacchino:2015hvk,Baek:2016lnv,Baek:2017ykw,Colucci:2018vxz,Biondini:2019int} that the annihilation among the VLQs themselves can 
give a significant contribution towards the relic density of the DM. 
The relic density of the DM with mass $m_S$ is given by \cite{McDonald:1993ex}:

\bea
\Omega h^2&=&\frac{1.09\times10^9 ~\rm{GeV^{-1}}}{g_*^{1/2}~M_{Pl}}\frac{1}{J(x_f)},
\eea
\label{relic_expression}
where $J(x_f)$ is given by 
\bea
J(x_f)&=& \int_{x_f}^{\infty}\frac{\langle \sigma |v|\rangle_{\rm eff}}{x^2}~\rm{dx}.
\label{th_int}
\eea

\noindent $\langle \sigma |v|\rangle_{\rm eff}$ in Eq.~\eqref{th_int} is the effective thermal average DM annihilation cross-sections including contributions from the co-annihilations and is given by 
\bea
\langle \sigma |v|\rangle_{\rm eff} &=& \frac{g_s^2}{g_{\rm eff}^2}\sigma(\overline{S}S)+2\frac{g_sg_{\psi}}{g_{\rm eff}^2}\sigma(\overline{S}\psi_i)(1+\Delta_i)^{3/2}~\rm{exp}[-\it{x}\Delta_i]\nonumber \\
&& +2\frac{g_{\psi}^2}{g_{\rm eff}^2}\sigma(\overline{\psi_i}\psi_j)(1+\Delta_i)^{3/2}(1+\Delta_j)^{3/2}~\rm{exp}[-\it{x}(\Delta_i+\Delta_j)]\nonumber \\
&& +\frac{g_{\psi}^2}{g_{\rm eff}^2}\sigma(\overline{\psi_i}\psi_i)(1+\Delta_i)^{3}~\rm{exp}[-2\it{x}\Delta_i].  
\label{eff_cs} 
\eea
\noindent In the equation above, $g_s$ and $g_{\psi}$ are the spin degrees of freedom for $S$ and $\psi_{i,j}$ where $\psi_{i,j}$ represents all the VLQs ($\mathcal{F}_1,\mathcal{F}_2,f$) involved. Here, $x=\frac{m_S}{T}$ and $\Delta_i$ depicts the
mass splitting ratio $\frac{m_i-m_S}{m_S}$, where $m_i$ stands for the masses of all the VLQs. $g_{\rm eff}$ in Eq. \eqref{eff_cs} is the effective degrees of freedom given by
\bea
g_{\rm eff} &=& g_s+g_{\psi}(1+\Delta_i)^{3/2}~\rm{exp}[-\it{x}\Delta_i].
\eea
Note that as the VLQs share the same $Z_2$ charge similar to the DM, their annihilations would also be important 
for evaluating the effective annihilation cross section. 
In the following analysis, we use the \texttt{MicrOmegas} package \cite{Barducci:2016pcb} to find the region of parameter space that corresponds 
to correct relic abundance for our DM candidate satisfying PLANCK constraints \cite{Aghanim:2018eyx}, 
\begin{equation}
\nonumber
0.119 \lesssim \Omega_{\rm{DM}} h^2 < 0.121.
\end{equation}

\begin{figure}[H]
\centering
\subfigure[]{
\includegraphics[scale=0.40]{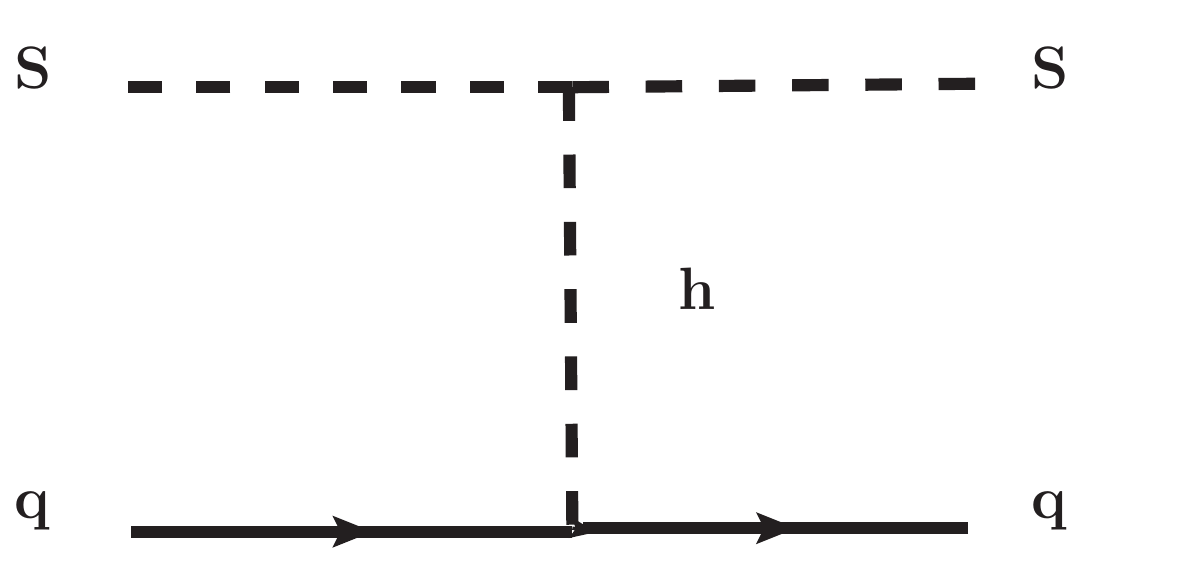}}
\subfigure[]{
\includegraphics[scale=0.40]{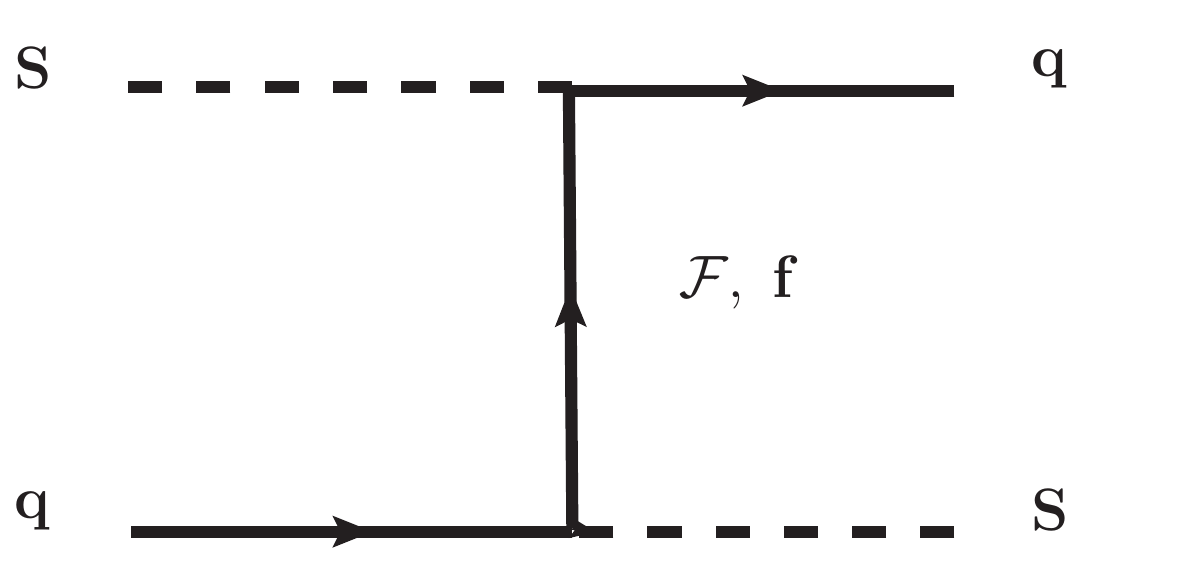}}
\subfigure[]{
\includegraphics[scale=0.40]{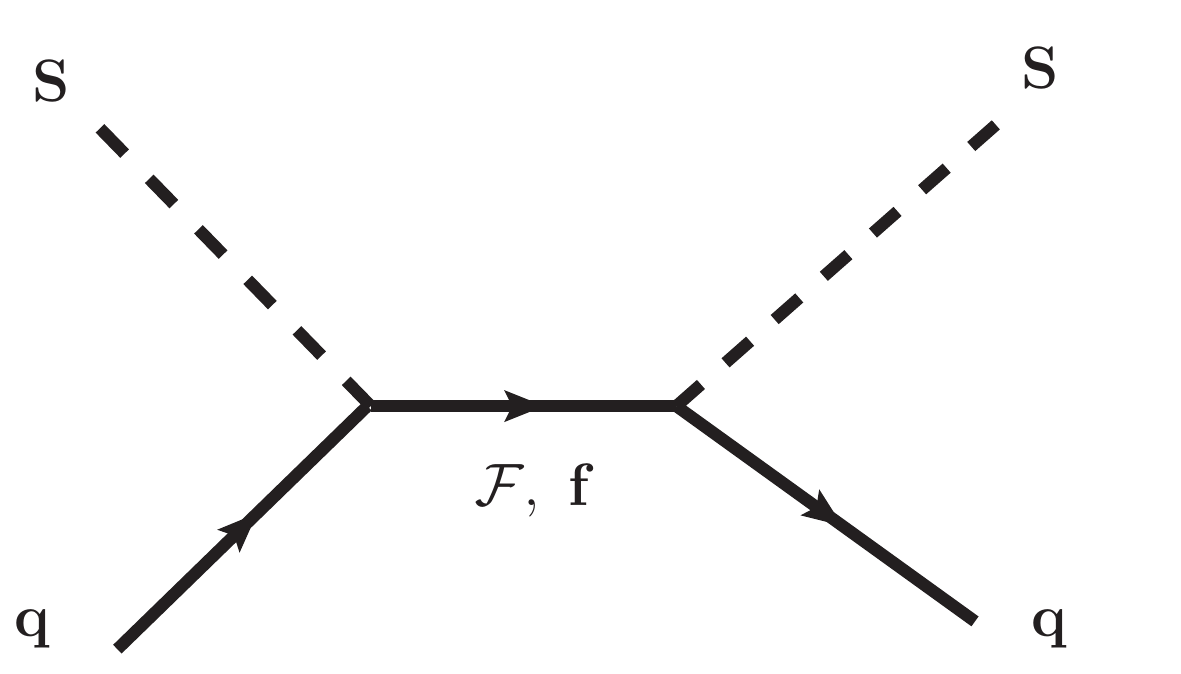}}
\caption{Spin independent elastic scattering of DM-nucleon.}
\label{DD}
\end{figure}
As mentioned earlier, the DM parameter space can be constrained significantly by the null result at different direct detection experiments such as LUX~\cite{Akerib:2016vxi}, PandaX-II~ \cite{Tan:2016zwf,Cui:2017nnn} and XENON1T~ \cite{Aprile:2017iyp,Aprile:2018dbl}. Apart from the usual SM Higgs mediated Feynman diagrams for the direct detection of the singlet scalar dark matter, the present setup has additional diagrams which come into the picture due to the presence of the VLQs. In Fig. \ref{DD} we show all the scattering processes of the DM $S$ with the detector nucleon. 

\subsection{Results}

Based upon the above discussion, it turns out that the following set of parameters are the relevant ones 
for DM phenomenology:
\begin{equation}
\{m_S, m_{1,2}, \sin\theta, \lambda_{HS}, \alpha_{1,2}\},
\end{equation}
whereas other parameters of the model $e.g.~ y$ and bare masses are derivable parameters from 
this set itself using Eq.~\eqref{dependent_parameters}. For simplicity, we set $\alpha_1 = \alpha_2 = \alpha$ unless otherwise 
mentioned. Since we specifically look for reopening the window of scalar singlet DM (which is otherwise 
ruled out) in the intermediate mass range $\sim (200-950)$ GeV through the co-annihilation process 
with the VLQs, we expect these VLQs to be heavier but having mass close to DM mass. However from the 
point of view of the LHC accessibility in future, it would be interesting to keep their masses lighter than 
1 TeV. With this, the value of $\sin\theta$ is expected to be small as seen from Eq.~\eqref{tan2theta}, unless there 
exists a very fine tuned mass difference $\Delta_{21}$ between the bare masses ($M_{\mathcal{F}}$ 
and $M_f$) of VLQs. As a benchmark value, we consider $\Delta_{21} = 50$ GeV. Note that with 
such a choice, the Yukawa coupling $y$, obtainable via relation Eq.~\eqref{dependent_parameters}(c) for fixed choices of $\Delta_{21}$ \
and $\sin\theta$, can be kept well below 1 provided $\sin\theta$ is small. We have taken here a conservative choice with $\sin\theta = 0.1$. Note that a small value of $y$ (significantly below 1) turns out to be quite 
natural from the point of view of maintaining perturbativity of $y$. It is also found in \cite{Gopalakrishna:2018uxn} that 
$y$ above 0.3 can make the Higgs quartic coupling $\lambda_H$ negative at some high scale and thereby 
could be dangerous for vacuum stability.

\begin{figure}[]
\centering
\subfigure[]{
\includegraphics[scale=0.39]{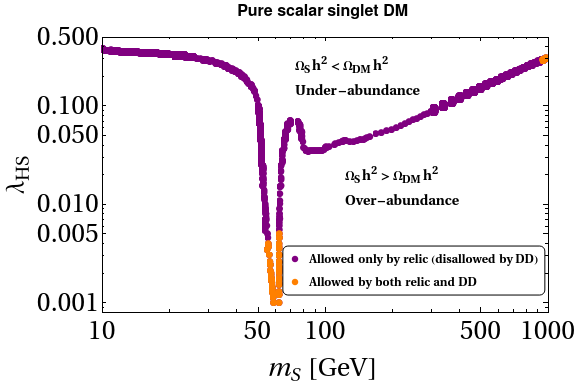}}
\subfigure[]{
\includegraphics[scale=0.39]{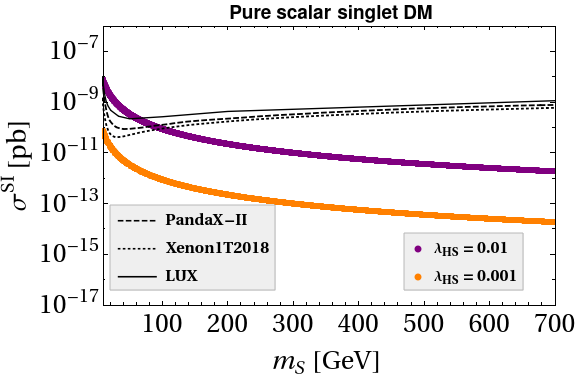}}
\caption{Relic density and direct search/XENON1T  allowed parameter space of the scalar singlet ($S$) DM on left (in $\l_{HS}-m_S$ plane). On right we show the spin independent DM-nucleon cross-section for a scalar singlet DM for two values of $\l_{HS}=0.01,0.001$.}
\label{pure_scalar_contour}
\end{figure}

To facilitate our discussion on DM parameter space of the model under consideration, we first provide the 
contour plot for correct relic density consistent with Planck data in the $\lambda_{HS} - m_S$ plane as 
shown in the left panel of Fig. \ref{pure_scalar_contour}. We also impose the DM-nucleon direct search cross section 
limit as obtained from XENON 1T experiment \cite{Aprile:2018dbl}. The relic as well as DD cross section satisfied points are indicated by the orange portion while points in purple portion of the contour indicates the relic-satisfied but otherwise excluded by DD limits. The change in the DD cross section versus $m_S$ variation with different choices of Higgs portal couplings are represented in the right panel. It turns out that singlet scalar DM mass below 1 TeV is essentially ruled out except in the SM Higgs resonance region. This finding is also consistent with the results of other works \cite{Bhattacharya:2019fgs,DuttaBanik:2020jrj}.

\begin{figure}[]
\centering
\subfigure[]{
\includegraphics[scale=0.39]{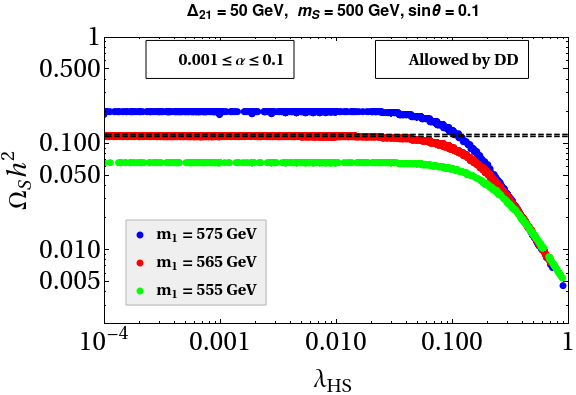}}
\subfigure[]{
\includegraphics[scale=0.39]{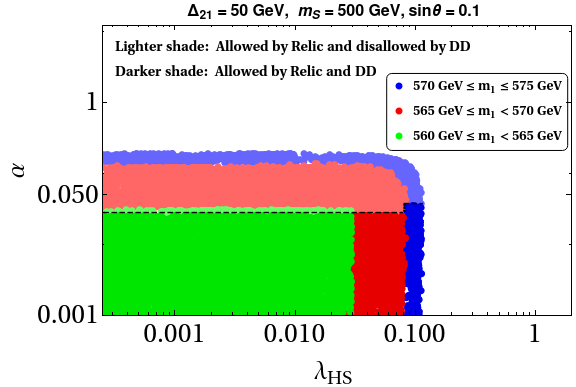}}
\caption{Left panel shows the variation of $\Omega_Sh^2$ with $\l_{HS}$ for different values of lightest VLQ mass ($m_1$) while right one describes allowed (disallowed) region for $\a-\l_{HS}$ parameters denoted by a darker (lighter) shade. The black dotted 
line (horizontal) in the left panel corresponds to the correct relic density. The black dotted line the right panel segregates the allowed 
region (below this line, having dark shaded points) from the dis-allowed region (above this line, having light shaded points) by direct 
detection results.}
\label{relic_500}
\end{figure}

In Fig. \ref{relic_500} left panel, we have shown the variation of relic density $\Omega_S h^2$ against the Higgs quartic coupling $\lambda_{HS}$ in our setup, while mass difference between the two up-type VLQs ($\Delta_{21}$), 
DM mass and mixing angle $\sin\theta$ are kept fixed at 50 GeV, 500 GeV and 0.1 respectively. Such a choice 
of DM mass is guided by our aim to explore the otherwise disallowed range of DM mass for scalar singlet DM,  
as state before. The respective plots corresponding to different $m_1$ masses are indicated by green (555 GeV), 
red (565 GeV) and blue (575 GeV) lines. The value of other remaining parameter $\alpha$ is varied in a range, 
$0.001 \leq \alpha \leq 0.1$, while generating the plots.

We find (refer to Fig. \ref{relic_500}, left panel) that the relic can be satisfied by relatively  
(compared to the pure singlet scalar case) small value of $\lambda_{HS} \lesssim 0.1$, thanks to the effect of 
co-annihilations (inclusive of diagrams in Fig. \ref{coann} and Fig. \ref{feynF}) involving the interactions provided in Eq. \eqref{lag} and the 
gauge interactions. The presence of this co-annihilation is more transparent when we observe that with the increase of $m_1$, the relic becomes larger. This is simply because with increased values of $m_1$, $\Delta_{1,2}$ become larger and hence the effect of co-annihilation reduces (see Eq. \eqref{eff_cs}) which contributes to a smaller effective annihilation cross section and hence larger relic is obtained.  
The small $\lambda_{HS}$ turns out to be helpful in evading the DD limits. Since some of the couplings involved in co-annihilation processes, $e.g.$ couplings $\alpha$, are also involved in the DD process, their effects become crucial in getting the correct DM relic density as well as to satisfy DD bounds. 
It can be noticed that for $\lambda_{HS}$ below 0.1, the effect of co-annihilation (actually the annihilations of VLQs) 
remains so important that 
the relic is essentially insensitive to the change of $\lambda_{HS}$. However this conclusion changes when 
$\lambda_{HS}$ crosses 0.1. There we notice a significant fall in the relic plots (with different $m_1$ value) as well as a merger of them. This is because in this regime, annihilation of DM provides a significant contribution to 
effective annihilation cross section $\langle \sigma v\rangle_{\rm eff}$ and co-annihilation, although present, becomes less important. So overall the decrease in relic beyond $\l_{HS} \gtrsim 0.1$ is effectively related to the increase in $\lambda_{HS}$ in the usual fashion. 

In the right panel of Fig. \ref{relic_500} we show the parameter space in the $\alpha-\l_{HS}$ plane, where the points with lighter shades are allowed only by the relic density and disallowed by the direct search experiments, whereas the darker shaded region (below the black dotted line separating allowed and the disallowed regions) corresponds to the parameter space which are allowed by both the relic density as well as the direct search constraints. In generating the plot, all other parameters except $m_1$ are kept at same values as in the plot of left panel. $m_1$ here is varied in three range of values as indicated by the color code mentioned in the inset of the figure. As with the increase of $m_1$ value (such that $\Delta_1$ also increases), the possibility of co-annihilation becomes less, the respective relic and DD satisfied region also shrinks. One can also notice that higher values of ${\alpha}$, beyond $\sim $0.05 or so, are disfavoured by the DD searches. The reason would be clear if we look the Feynman diagrams in Fig. \ref{coann} and \ref{DD}. A large $\alpha$ would lead to a cross section relatively large compared to the DD limits. Hence it turns out that in the regime of parameters with small values for both $\lambda_{HS}$ and $\alpha$, it is the VLQ annihilations via gauge interactions (as part of co-annihilations via Fig. \ref{feynF}) which play crucial role in getting the correct relic density. 

\begin{figure}[H]
\centering
\includegraphics[scale=0.39]{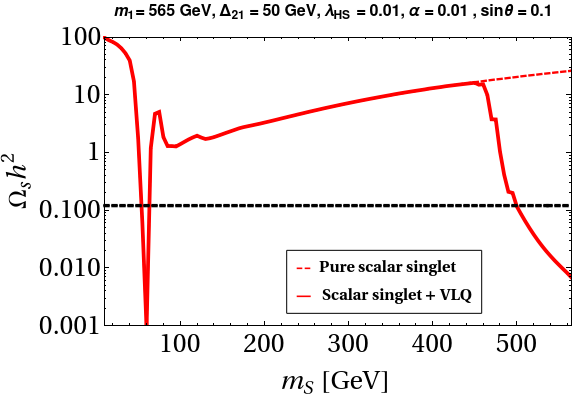}
\caption{Relic density plot comparing the scenario for scalar singlet DM with that of scalar singlet DM in presence of vector like quarks with $\l_{HS} = 0.01$.}
\label{relic-our-case}
\end{figure}
A variation of relic $\Omega_S h^2$ with mass of the DM $m_S$ is shown in Fig. \ref{relic-our-case} indicated 
by red line where $m_1, \lambda_{HS}, \Delta_{21}, \alpha$ and $\sin\theta$ are fixed at 565 GeV, 0.01, 50 GeV, 0.01 and 0.1 respectively as also mentioned on top of the figure. While compared with the case of a pure singlet DM having Higgs portal coupling set at 0.01 (the dashed red line), we notice several interesting features. Firstly the pattern  is very much similar till $m_S$ approaches 450 GeV or so. Beyond this point, the co-annihilations 
starts to be dominant as the $\Delta_1$ becomes small as seen from Eq. \eqref{eff_cs}. Hence due to the sudden increase in the effective-annihilation cross section, the relic abundance falls rapidly. With $m_1$ value fixed at 565 GeV, $m_S$ can not exceed $m_1$.

Below in Fig. \ref{DD_alpha} left panel, we provide the DD cross section versus $m_S$ in our setup for two different choices of $\alpha$. While $\alpha$ = 0.01 is allowed for most of the intermediate mass range of DM, 
$\alpha$ = 0.05 remains above the DD limits set by XENON 1T data, in accordance with the contribution followed from Fig. \ref{relic_500}. Also there is a noticeable dip in the plot which is indicative of the relative sign difference between these $s$ and $t$ channel contributions.  In the right panel, we explore what happens if we deviate from our simplified assumption of keeping both $\alpha_1$ and $\alpha_2$ same. We find that in this case $\alpha_1$ = 0.01 and $\alpha_2$ = 0.03 are also allowed set of choices. However in terms of usefulness of VLQs in bringing back the otherwise disallowed intermediate mass range of a singlet scalar DM, our conclusion remains unaltered. In the following section we extend our discussion for EW vacuum stability and in the next, we find the parameter space consistent with both the DM phenomenology as well as vacuum stability.   

\begin{figure}[H]
\centering
\subfigure[]{
\includegraphics[scale=0.39]{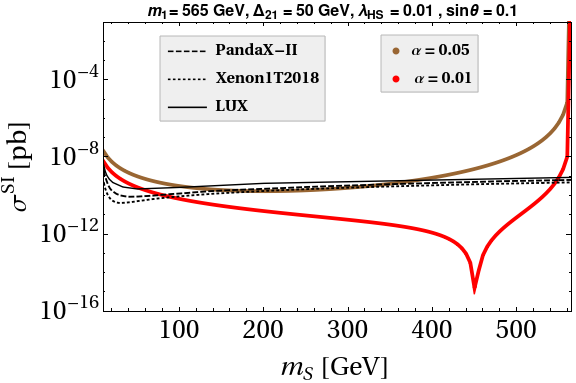}}
\subfigure[]{
\includegraphics[scale=0.39]{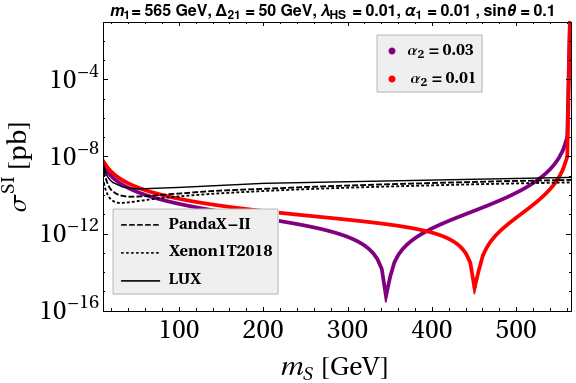}}
\caption{Spin independent direct detection cross-section for : (a) two different values of $\a=0.05,0.01$ and (b) fixed value of $\a_1=0.01$ and two different values of $\a_2=0.03,0.01$} 
\label{DD_alpha}
\end{figure}

\section{Electroweak Vacuum Stability}
\label{sec:highscale}

In this section, we aim to discuss the impact of the inclusion of the singlet scalar and VLQs on 
EW vacuum stability in our setup. It is well known that the absolute stability of the SM Higgs 
vacuum can be ensured by  $\lambda_H (\mu) > 0$ at any energy scale $\mu$ provided the 
EW minimum is the global one. In presence of an additional deeper minimum other than the 
EW one, one has to evaluate the tunnelling probability $P_T$ of the EW vacuum to this second 
minimum given by $P_T = T_U^4 \mu_B^4 e^{-8\pi^2/(3 |\lambda_H (\mu_B)|)}$. Here $T_U$ 
is the age of the Universe, $\mu_B$ is the scale at which probability is 
maximised, to be obtained from $\beta_{\lambda_H} (\mu_B) = 0$ where $\beta_{\lambda_H}$ is the beta function for SM Higgs quartic coupling. The Universe would then 
be a metastable one if the decay lifetime of the EW vacuum to the second one is longer than 
the age of the Universe, $T_U$ . This metastability requires 
\begin{equation}
\lambda_H (\mu_B) > \frac{-0.065}{1-0.01 {\rm{ln}}\left( \frac{v}{\mu_B}\right)}.
\end{equation}

Within the SM alone, it is the top quark Yukawa coupling, $y_t$, which drives the SM Higgs quartic coupling 
to negative value at around $10^{10}$ GeV \cite{Buttazzo:2013uya,Tang:2013bz,Ellis:2009tp,EliasMiro:2011aa,Degrassi:2012ry}. However within the present limits on the top quark mass, the 
EW vacuum turns out to be a metastable one. In our setup, inclusion of singlet scalar $S$ as DM and the 
VLQs would affect this situation. The presence of an additional singlet scalar is known to generate a 
positive contribution to the beta function of $\lambda_H$ through the interaction like $\frac{\l_{HS}}{2} 
H^{\dagger}HS^2$ and thereby helps in driving Higgs vacuum toward more stability \cite{Haba:2013lga,Khan:2014kba,Khoze:2014xha,Gonderinger:2009jp,Gonderinger:2012rd,Chao:2012mx,Gabrielli:2013hma,DuttaBanik:2018emv,Ghosh:2017fmr}. It is found that in order to 
make the Higgs vacuum absolutely stable upto $M_{\rm Pl}$, the scalar singlet (here DM) mass should fall 
above $\sim$ TeV. As in this paper, we focus on the intermediate range of DM mass, its sole presence 
cannot be effective in achieving the absolute vacuum stability. On the other hand, we also have the VLQs. 
It turns out, as discussed below, their presence would be helpful in achieving the absolute stability of the EW vacuum even with the scalar singlet DM mass below 1 TeV.

We have already seen the important role of these VLQs in the present set up in making the otherwise 
disallowed parameter space (say in terms of mass) a viable one by providing additional contributions to 
DM effective annihilation cross section and in DD diagrams. Following the results of 
\cite{Gopalakrishna:2018uxn}, these VLQs 
also help in achieving the electroweak vacuum stable up to a large scale. This can be understood if we 
look into their positive (additional) contribution to the beta function of the gauge couplings $g_{1,2,3}$ associated 
with $SU(3), SU(2)$ and $U(1)$ gauge group of the SM respectively (at one loop) given by 
\besub
\bea
\b{g_1} &=&  \b_{g_1}^{\rm{SM}} +\b_{g_1}^{\rm{VLQ}}~ = ~ \b_{g_1}^{\rm{SM}} + \frac{g_1^3}{16 \pi^2}\bigg{[}\frac{4}{5}N_c~(2n_2Y_{\mathcal{F}}^2+n_1Y_{f^{\prime}}^2)\bigg{]}  \\
\b{g_2} &=&  \b_{g_2}^{\rm{SM}} +\b_{g_2}^{\rm{VLQ}}~ = ~\b_{g_2}^{\rm{SM}}+ \frac{g_2^3}{16 \pi^2}\bigg{(}\frac{2}{3}N_c~n_2\bigg{)}\\
\b{g_3} &=&  \b_{g_3}^{\rm{SM}} +\b_{g_3}^{\rm{VLQ}}~ = ~\b_{g_3}^{\rm{SM}} + \frac{g_3^3}{16 \pi^2}\bigg{(}\frac{2n_3}{3}\bigg{)}
\eea
\label{beta_g1g2}
\eesub
where colour charge $N_c=3$ for any fermion (SM or VLQ) belonging to the fundamental representation of $SU(3)$, $n_1$ and $n_2$ represent number of $SU(2)$ singlet and doublet VLQs respectively while $n_3$ corresponds to the number of $SU(3)$ triplet vector like fields. $Y_{f^{\prime}}$ and $Y_{\mathcal{F}}$ are the hypercharges of the VLQ  singlets and doublets respectively.  The VLQs being charged under $SU(3)$ increases the number of coloured particles in the present setup and hence effectively increases the 
$\b$-function of the gauge couplings, noticeably for the the $g_3$. 
As a result of this, the running of the top Yukawa coupling $y_t$ experiences a sizeable decrease in its value (compared to its value within SM alone) at high scale predominantly due to the 
involvement of $-8y_t g_3^2$ term in $\b^{SM}_{y_t}$ as 
\bea
\b{y_t} &=&\beta_{y_t}^{SM}+\beta_{y_t}^{VLQ} = \frac{y_t}{16\pi^2} \bigg{[} \frac{9}{2}y_t^2 +2N_c y^2 - \frac{17}{20} g_1^2 - \frac{9}{4} g_2^2 -8g_3^2+2N_c y^2 \bigg{]}.
\label{run-yt}
\eea
\noindent Provided this 
decrease in $y_t$ is significant enough, it may no longer drag the $\l_{H}$ towards the negative value (recall 
that it was the $-6y_t^4$ term present in $\b_{\l_H}$ which pushes $\l_H$ to negative in SM at around $10^{9-10}$ GeV ($\equiv \Lambda^{SM}_I$) depending on the top quark mass), 
rather keeps it positive all the way till the Planck scale. 

One should also note that due to the involvement of both doublet as well as singlet VLQs in our framework, 
Yukawa interaction involving the SM Higgs and these VLQs will also come into the picture (see Eq.~\eqref{lag}). 
The associated coupling $y$ involved in this interaction may result negative contribution in the running of $\l_H$
as seen from, 
\bea
\b{\l_H} &=&  \b_{\l_H}^{\rm{SM}} +\b_{\l_H}^{\rm{VLQ}}~ + \b_{\l_H}^{S(1)}, \nonumber \\
&=& ~\b_{\l_H}^{\rm{SM}} + \frac{2n_F}{16 \pi^2}~(4N_cy\l_H-2N_cy^4) +\frac{1}{2}\l_{HS}^2,
\label{run-LH}
\eea
where $n_F$ is the number of families of VLQ multiplets (both doublet and singlet) participating in the Yukawa interaction with SM Higgs. 
In fact, it has been shown in \cite{Gopalakrishna:2018uxn} that $y>\mathcal{O}$(0.3) can make $\l_H$ negative 
at the large scale. However in our analysis we have found that this coupling does not have a significant impact on the DM phenomenology and hence we can keep it rather small throughout. Note that this smallness is in fact supported 
by the small mixing angle $\theta$ and $\Delta_{21} \sim \mathcal{O}(100)$ GeV as can be seen using Eq. \eqref{dependent_parameters}. 

For the analysis purpose in both the VLQ as well as the VLL scenarios, we run the two-loops RG equations for all the SM couplings as well as all the other relevant BSM coupling involved in both the setups from $\mu = m_t$ to $M_{\rm Pl}$ energy scale. We use the initial boundary values of all SM couplings as given in table \ref{initial_conditions} at an energy scale $\mu=m_t$ . The boundary values have been evaluated in \cite{Buttazzo:2013uya} by taking various threshold corrections at $m_t$ and the mismatch between top pole mass and $\overline{\rm MS}$ renormalised couplings into account. Here, we consider $m_h=125.09$ GeV, $m_t=173.2$ GeV, and $\a_S(m_Z)= 0.1184$. 
We only provide the one-loop $\b$-functions for both the scenarios namely, VLQ and VLL in appendix \ref{RGE_VLQ} and appendix \ref{RGE_VLL} respectively. 
The $\b-$functions were generated using the model implementation in \texttt{SARAH} \cite{Staub:2013tta}. 
\begin{table}[H]
\centering
\begin{tabular}{|c|c| c| c | c|c| c| c|c|}
\hline
Scale & $\l_{H} $  & ~$y_{t}$ &  ~$g_{1}$& ~$g_{2}$& ~$g_3$\\  
\hline
$\mu=m_t$ &$0.125932$ & $0.93610$ & $0.357606$ & $0.648216$ & $1.16655$  \\
 \hline
\end{tabular}
\caption{Values of the relevant SM couplings (top-quark Yukawa $y_t$ , gauge couplings $g_i (i = 1, 2, 3)$ and Higgs quartic
coupling $\l_H$ ) at energy scale $\mu= m_t= 173.2$ GeV with $m_h =125.09$ GeV and $\a_S(m_Z)= 0.1184$.
 }
\label{initial_conditions}
\end{table}
\noindent 

In continuity with the discussion above, we now provide plots for the running of the gauge couplings, Yukawa coupling $y_t$ and Higgs quartic coupling $\lambda_H$. For this purpose, we consider the set of benchmark values of parameters of our 
setup denoted by BP I in table \ref{BP I}, the choice of which are mainly motivated from the DM phenomenology discussed before.

\begin{figure}[H]
\centering
\subfigure[]{
\includegraphics[scale=0.39]{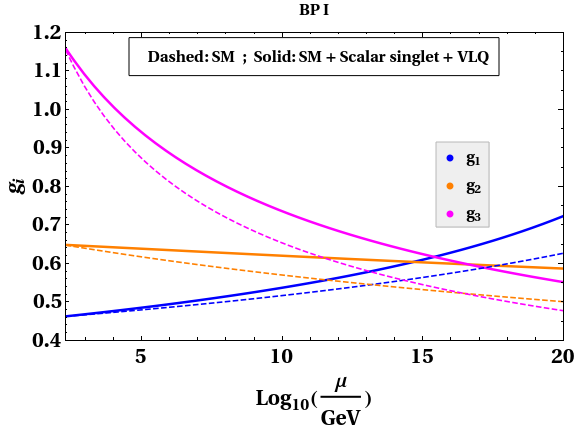}}
\subfigure[]{
\includegraphics[scale=0.4]{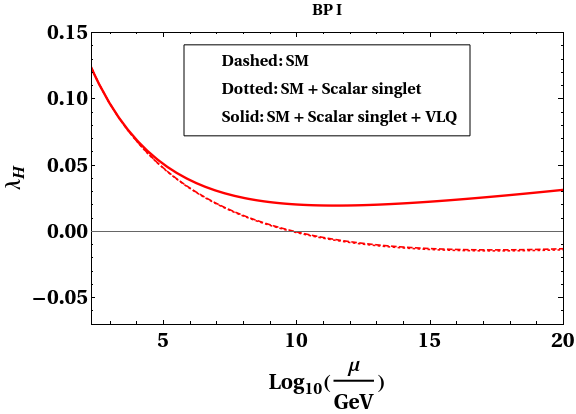}}
\caption{Running of (a) gauge couplings and (b) Higgs quartic coupling.}
\label{running-a}
\end{figure}

\begin{table}[H]
\centering
\begin{tabular}{|c|c| c| c | c|c| c| c|c|}
\hline
BP & $m_{S} ~\rm{[GeV]}$  & ~$m_1 ~\rm{[GeV]}$ & ~$\Delta_{21} ~\rm{[GeV]}$& ~$\lambda_{HS}$& ~$\a$& ~$\Omega_{S}h^2$ &$\sigma_{S}^{SI}$($pb$)\\  
\hline
BP I &$500$ & $565$ & $50$ & $0.01$ & $0.01$ & 0.119& $7.169\times10^{-12}$ \\
 \hline
\end{tabular}
\caption{Benchmark point which satisfy the correct relic density and are allowed by the direct detection experiments for $\sin{\theta}=0.1$.} 
\label{BP I}
\end{table}
In the last two columns, we specify the respective contribution to the relic and DD cross section estimate followed from this particular choice of parameters, BP I. Using this, we first show the effect of VLQs on gauge couplings $g_1$ (in blue), $g_2$ (in orange) and $g_3$ (in purple) in the left panel of Fig. \ref{running-a}. While their running 
in SM is represented in dotted lines, the respective running of them in case of the present model (SM+VLQs+scalar singlet) are denoted by the solid lines. 
While the positive shift to $g_3$ is expected due to the inclusion of VLQs, the rise in $g_1, g_2$ values 
is due to the presence of the extra $SU(2)$ doublet i.e. $\mathcal{F}$ in our framework.
In Fig. \ref{running-a} (b) we compare 
the running of Higgs quartic coupling in our model (solid line) with that of the SM (dashed line) 
\begin{figure}[H]
\centering
\subfigure[]{
\includegraphics[scale=0.39]{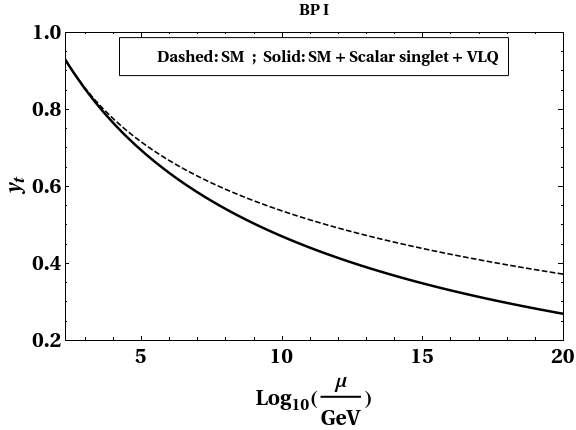}}
\caption{Running of top Yukawa coupling.}
\label{yukawa1}
\end{figure}
\noindent and scalar singlet extension of the SM (dotted line). Note that with the choice of $\l_{HS} = 0.01$, 
introduction of the scalar singlet does not make any significant contribution on the SM prediction of $\b$-function. As we have discussed above, the presence of VLQs alter the running of the gauge couplings with a positive shift and hence 
top quark Yukawa, $y_t$, has a prominent drop at high scale. In Fig. \ref{yukawa1}, this downward shift in the 
running of top Yukawa coupling (due to the negative contribution coming from the gauge coupling $g_3$ and 
others) is shown.

\section{Combined analysis of DM and EW vacuum stability}
\label{DM_VS}

\begin{figure}[H]
\centering
\includegraphics[scale=0.5]{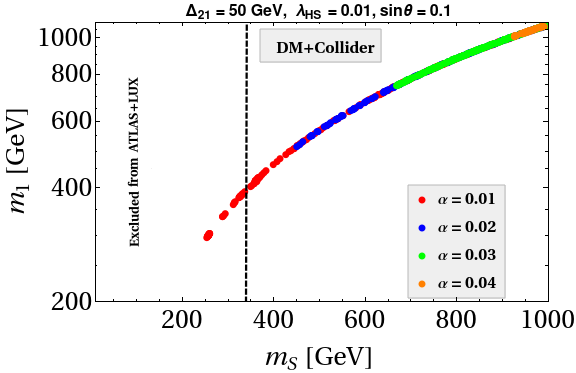}
\caption{Allowed parameter space in $m_1-m_{S} $ plane. }
\label{m1ms}
\end{figure}

Fig. \ref{m1ms} shows the allowed parameter space in the $m_1 - m_S$ plane where the dark matter relic and DD constraints are imposed. In obtaining this plot, parameters other than $m_1, m_S$ and $\alpha$ are kept fixed at 
their respective benchmark values (BP I of table \ref{BP I}). Note that the patch with red dots associated to 
$\alpha = 0.01$ spans over the entire mass range of dark matter $m_S: 250$ GeV to 1 TeV while a slight increase 
in $\alpha$ (= 0.02) restricts $m_S$ to be above $\sim$ 450 GeV. Similar trend follows for larger values of $\alpha 
= 0.03 (0.04)$ for which the allowed DM mass starts from 650 GeV (925 GeV). This happens due to combined 
impact of (i) the involvement of the mass splitting $ m_1 - m_S$ in determining the relic density and (ii) sensitivity of the dark matter DD to $\alpha$. The correlation between $m_1$ and $m_S$ as seen in Fig. \ref{m1ms} follows mainly 
due to the relic density requirement co-annihilations of VLQs to SM particles are the relevant processes that bring 
the relic density of scalar singlet DM at an appropriate level. As we have discussed before, this depends crucially 
on the mass splitting of $m_1 - m_S$. However as it turned out from our understanding of DM phenomenology, 
relic density is almost insensitive to $\alpha$ though the DD cross section depends on it. Hence an increase in 
$\alpha$ value would also enhance the DD cross-section (thereby excluded) unless the mass of the DM is also increased (as cross section varies inversely with $m_s$). This explains why with the increase of $\alpha$, the 
DM mass should also be increased (as seen going from red $\rightarrow$ blue $\rightarrow$ green and then to 
orange patches in the plot). 

Now turning back to vacuum stability issue, we note that presence of VLQs allows the EW minimum to be absolutely stable all the way up to $M_{\rm Pl}$. However this conclusion is independent to the choice of $\alpha$ or in other words, $\alpha$ remains unrestricted. In our setup, we have Yukawa interaction of the VLQs to the DM, guided by $\alpha$. Now with the inclusion of DM phenomenology, we restrict the coupling $\alpha$. In conclusion, we have ended up 
in a situation, where a scalar singlet DM having intermediate mass range (in particular $\sim$ 250-900 GeV) now becomes allowed in presence of VLQs and simultaneously can make the EW stable. 


In this scenario with VLQs, as we have seen, there arises several new annihilation and co-annihilation channels 
due to the additional interactions of DM present in Eq. \eqref{lag}. These are shown in terms of Feynman diagrams 
in Fig. \ref{feynS} and \ref{coann}. Due to the presence of the last two terms in the Eq. \eqref{lag}, one can see 
that the dark matter $S$ can also annihilate into gluon pair ($gg$) at one loop (through box diagram). It has been shown in \cite{Giacchino:2015hvk,Baek:2016lnv,Baek:2017ykw,Colucci:2018vxz,Biondini:2019int} that with larger choice of Yukawa coupling (here $\alpha$) this annihilation of $S$ to the gluon pair can provide a significant contribution to the relic of $S$. In \cite{Baek:2016lnv}, the authors notice that for the mass regime $m_S<m_t$ 
with the Yukawa coupling of the $\sim \mathcal{O}(10)$ the correct relic density can be satisfied. This large 
Yukawa coupling (involving up-quark) however can create a problem in the DD of the dark matter as it will 
generate a large spin independent cross-section and hence can be in the verge of being excluded by the DD searches. One should 
note that the amplitude of the $SS\rightarrow gg$ depends on the square of the product of this Yukawa coupling 
and the strong gauge coupling. Hence this process will dominate over the other annihilation or the co-annihilation channels of $S$ only if the associated Yukawa coupling is large. In our scenario, first of all, we focus 
on the scalar DM having mass above the top threshold. Secondly, we have used a small Yukawa in general, 
$i.e. ~\alpha \sim 0.01$ which helps in evading the DD constraint. Hence $SS\rightarrow gg$ (and other 
processes like $ss\rightarrow \gamma \gamma$, three body final states $etc$.) remains suppressed in our case.

We find that the annihilations of VLQs take part significantly in the effective DM annihilation cross section in 
this scenario. As shown in earlier works \cite{deSimone:2014pda,Giacchino:2015hvk,ElHedri:2017nny,Colucci:2018vxz}, such annihilations of the two VLQ particles, in the non-relativistic limit, can be affected by the 
non-perturbative Sommerfeld corrections through gluonic exchanges. It has been noticed that this correction 
may affect the relic abundance calculation by at most $15\%$ (see \cite{Giacchino:2015hvk,Colucci:2018vxz} 
for a recent discussion and references therein). In the present set up we have not included this correction, we 
leave it for a more complete exploration as a future study. One should also take into account 
the possible formation of the bound states due to the presence of the VLQs. It has however been concluded 
\cite{Mitridate:2017izz,Biondini:2018pwp}, using the set up similar to ours except that only SM singlets are involved there, that the bound state have only moderate impact on the relic density and hence negligible. 

It might be pertinent here to discuss the possible signatures of these additional colored particles in colliders. First note 
that lightest VLQ decaying into SM quark and $W$ boson leading to a prominent signature through jets + $\slashed E_T$ 
+ charged lepton \cite {Buckley:2020wzk} is forbidden in our set-up as an artifact of $Z_2$ symmetry (for which the 
direct Yukawa interaction involving VLQs, higgs and SM quarks is absent). Secondly, the possibility to trace back VLQ's 
signature through its decay into top quark and dark matter leading to $\slashed E_T$+ hard jets also becomes 
non-existent due to the unavailability of the phase space with the choice of parameters dictated by the DM phenomenology. 
The mass-spitting between $m_S$ and $m_1$ turns out to be $\sim \mathcal{O}$(50) GeV as seen\footnote {The heaviest 
VLQ's mass ($m_2$) is not that much restricted from the DM phenomenology or from vacuum 
stability point of view as long as $\sin\theta$ remains small. Only for representative purpose, we consider $m_2$ to be 
heavier than $m_1$ by 50 GeV and eventually $m_2$ - $m_S$ remains still below $m_t$. The other VLQ's mass 
(we have a $SU(2)_L$ doublet and a singlet VLQs), $\mathcal{M_F}$, falls in between $m_1$ and $m_2$ and plays 
important role in co-annihilations among VLQs.} from Fig. \ref{m1ms}. As we have discussed in section \ref{constraints}, 
presence of these VLQs in the current setup can only be felt in collider through their subsequent decay into the DM 
particle $S$ and light SM quark (or antiquark). Hence missing energy ($\slashed E_T$) plus multi-jets (arising due to gluon emission) would be the signal at LHC. Though the study in \cite{Giacchino:2015hvk} shows that a VLQ mass $\sim$ 550-750 
GeV is allowed with $m_S \sim 500$ GeV, such a constraint too won't be directly applicable here. This constraint came primarily 
due to the presence of large Yukawa coupling between VLQs, DM and the SM quark required to satisfy the DM relic density in 
their scenario. In our case, the correct relic is effectively produced by the annihilation of VLQs and the Yukawa coupling 
$\alpha$ is restricted to be small (to satisfy the DD bound). In fact, $\alpha$ is kept well below 0.1 in our case (see Fig. \ref{m1ms}). Still in Fig. \ref{m1ms} we apply $m_1 > $ 400 GeV and $m_S>$ 300 GeV (discussed in \cite{Giacchino:2015hvk}) as a conservative bound (indicated by the dotted vertical line). In the present work, we do not perform a detailed collider study to obtain precise bounds on mass of the VLQs. However 
it seems that the study would be rich due to the presence of both doublet and singlet VLQs and can surely be taken as 
a future endeavour.

\section{Scalar singlet DM with vector like leptons}
\label{sec:vll}

In this section, we investigate a similar setup where vector like quarks are replaced with vector like leptons. 
While we expect some similarities with respect to DM phenomenology, the sudy of vacuum stability will be different. 
For this purpose, we consider the extension of the SM by a scalar singlet DM $S$ along with a vector like lepton doublet $E = (E^0, E^-)^T$ and a neutral singlet lepton $\chi$, all of them being odd under the additional $Z_2$ 
while SM fields are even under the same. The Lagrangian involving the VLLs can then be written as follows : 
\bea
-\mathcal{L}^{\rm VLL} = y_{l_{ij}} \bar{E_{i}} \tilde{H}\chi_{j} + \a_{l_{i\b}} \bar{E}_{R_i} S l_{L_\b}    + (M_E)_{ij} \bar{E}_i E_j + (M_{\chi})_{ij} \bar{\chi}_i \chi_j+ h.c,
\label{lagvll} 
\eea
\noindent where $i,j$ are generation indices associated with the additional vector like lepton doublets and singlets 
and $\b=e,\m,\tau$. We will specify the required number of generations whenever applicable. Here $E_i$ represents the $SU(2)_L$ VLL doublets with hypercharge $Y=-\frac{1}{2}$ whereas $\chi_i$ represents the $SU(2)_L$ VLL singlets with hypercharge $Y=0$. We now proceed with only one generation of VLL doublet 
and singlet. Thereafter, we discuss implications of presence of more generations. 

After the electroweak symmetry breaking, there will be a mixing among the neutral components of the added VLL doublet and singlet originated due to the presence of the first term in the above Lagrangian. This is similar to 
the mixing $\theta$ of section \ref{model}, and here we denote it by $\theta_l$ defined by $\tan 2\theta_l = \sqrt{2} y_l 
v/(M_{\chi} - {M_E})$, analogous to Eq. \eqref{tan2theta}. The mass eigenvalues $m_{1,2}$ follow the same relation as 
provided in Eqs. \eqref{vlqmass}, replacing $\theta$ by $\theta_l$, $\mathcal{F}$ by $E$ and $f$ by $\chi$. 
In the scalar sector, the relations Eqs. \eqref{p1}-\eqref{sclrmass} remain unaltered. 

Since here we will study the scalar singlet DM and vacuum stability in presence of the VLLs,  constraints like 
stability, perturbativity, relic and DD limits as mentioned in section \ref{constraints} are also applicable in this 
study.  Additionally one should consider the constraints coming from the collider searches of the VLLs. The 
limit from the LEP excludes a singly charged fermion having mass below 100 GeV \cite{Achard:2001qw,Abdallah:2003xe}, as a result we consider $m_{E^{\pm}}\geq 100$ GeV. Due to the presence of the charged VLL, the 
present setup can also be probed at the LHC. Being a part of the $SU(2)$ doublets, the charged VLLs can be 
pair-produced at the LHC in Drell-Yan processes, $pp \rightarrow $ vector like leptons and anti-leptons pair, and subsequently undergo Yukawa-driven decays into a DM scalar and a charged SM leptons, $E^{\pm} \rightarrow S l^{\pm}$ and thus leading to a characteristic opposite sign di-lepton plus missing energy ($\slashed{E_T}$) signature \cite{Kowalska:2017iqv,Chakraborti:2019fnz}. In \cite{Bhattacharya:2017sml,Bhattacharya:2016rqj} 
which also has a similar setup as ours, the authors have shown that if the mass splitting between the charged and the neutral VLL is
less than mass of $W^{\pm}$ boson, then $E^{\pm}$  can also decay via three body suppressed process: $E^{\pm} \rightarrow E^{0}l^{\pm}\nu$ giving rise to the displaced vertex signature at LHC. 

The additional VLLs in the present setup can also contribute to the electroweak precision test parameters 
$S, T$ and $U$ \cite{Peskin:1991sw,delAguila:2008pw,Erler:2010sk}. The values of these parameters are tightly constrained by experiments. As discussed above, due the presence of doublet and a singlet VLL and their interaction with the SM Higgs (for one generation of VLLs), after the electroweak symmetry breaking three physical states namely $i.e.$ one charged and two neutral are obtained. Therefore, the contribution to the precision parameters depends on the masses of the physical states as well as on the mixing angle $\sin{\theta_l}$ in the present setup. One can look the study in \cite{Bhattacharya:2018fus} where a detailed analysis of electroweak precision test parameters $S, T$ and $U$ for the setup similar to ours has been presented. 

\subsection{DM phenomenology in presence of VLL}
\label{DM_VLL}

DM annihilations will proceed through the diagrams similar to the ones mentioned in Fig. \ref{feynS},       \ref{coann} and \ref{feynF}, replacing the VLQ 
by their VLL counterpart. However as the VLLs do not carry any colour quantum number, gluon involved processes would not be present. Other gauge mediated diagrams are present though and continue to play important roles as we will see below.  

\begin{figure}[H]
\centering
\subfigure[]{
\includegraphics[scale=0.39]{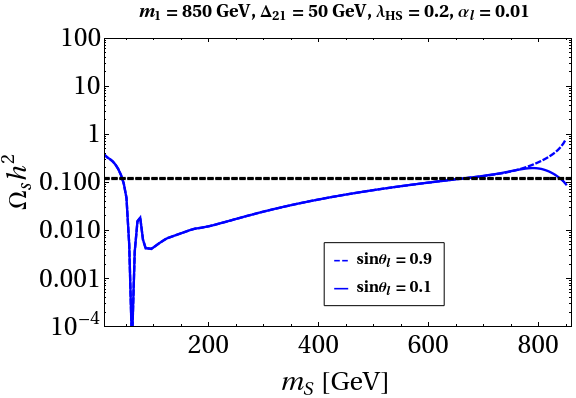}}
\subfigure[]{
\includegraphics[scale=0.39]{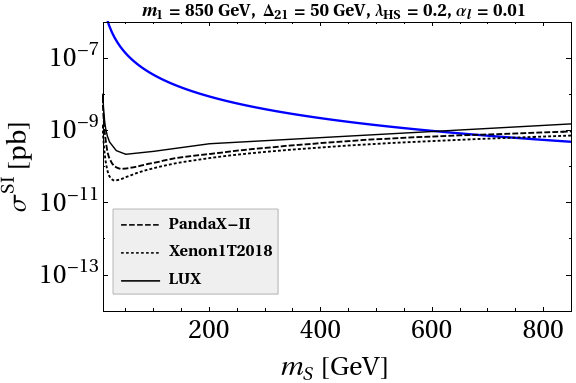}}
\caption{Relic density and direct search/XENON1T  allowed parameter space of the scalar singlet ($S$) DM + VLL for different $\sin\theta_l$.} 
\label{VLL-relic-dd-plots}
\end{figure}
We provide in the left panel of Fig. \ref{VLL-relic-dd-plots} the DM relic density ($\Omega_S h^2$) versus mass of the DM ($m_S$) plot, while parameters like $\Delta_{21}$ (= $m_2 - m_1$) and $\alpha_l$ (assuming $\a_{l_{i\b}}=\alpha_l \delta_{i\b}$) are fixed at 50 GeV and 0.01 respectively, same as their respective 
values in case of VLQs. However Higgs portal coupling of the DM is kept fixed at an enhanced value, $\l_{HS} = 0.2$ 
as compared to its value in case of VLQ. The  requirement of a larger $\l_{HS}$ would be evident when we 
discuss the vacuum stability. Now with such a large $\l_{HS}$, the pure singlet scalar DM relic density would be under-abundant for DM mass below 600 GeV as evident from left panel of Fig. \ref{pure_scalar_contour}. Furthermore presence of VLL having mass suitable for any possible co-annihilation would make the situation 
worsen. Hence we consider a relatively heavy VLL mass set at 850 GeV. Note that there is a possibility of satisfying relic by a $\sim$ 660 GeV DM where no co-annihilation involving VLL takes place. Although the relic is satisfied, this 
particular $m_S = 660$ GeV with $\l_{HS}$ = 0.2 (the very first intersection with correct relic line indicted 
by the horizontal line) is ruled out due to the associated large DD cross section, unless a heavier DM is considered 
as seen from right panel of Fig. \ref{VLL-relic-dd-plots}.

The VLLs start to contribute through co-annihilation around $m_S \sim$ 790 GeV and above as indicated 
by the drop of the relic curve associated with $\sin\theta_l = 0.1$ in the left panel of Fig. \ref{VLL-relic-dd-plots} 
which finally satisfies the relic at $m_S = 840$ GeV. This pattern is similar to the one we have obtained in case 
with VLQs. Here annihilations of VLLs through gauge mediation play the significant role (almost 80 percent contribution to the relic density of $S$). Another interesting pattern is observed for a large $\sin\theta_l$. As 
seen in the left panel of Fig. \ref{VLL-relic-dd-plots}, there is no such change observed with $\sin\theta_l = 0.9$. 
It mainly follows the pattern of a regular scalar singlet relic contour plot. Note that with a large $\sin\theta_l \sim 0.9$,  
the lightest VLL  with mass $m_1$ taking part in the co-annihilation and annihilation of VLLs to SM particles 
consists of mostly the singlet component $\chi$ as compared to its dominant doublet nature in case with 
small $\sin\theta_l$. Hence gauge mediated processes are naturally suppressed and there is no such effective annihilation of VLL). With large $\theta_l$, the only possibility where DM can satisfy the relic and DD cross section is with large $\alpha_l$. As the DD cross section involves only the Higgs mediated contribution similar to the one 
shown in Fig. \ref{DD}  (which only depends on $\l_{HS}$), the plot in right panel is essentially independent of $\sin\theta_l$ and $\alpha_l$.

\subsection{Vacuum stability in presence of VLL}
\label{EWV_VLL} 

With the above understanding about the relevant parameter space from DM phenomenology, we are 
now going to analyse the impact of the VLL + scalar singlet set up on vacuum stability. For this purpose, we 
employ the new set of RG equations provided in the appendix \ref{RGE_VLL}.  In doing this 
analysis, we consider more than one generation of doublet-singlet VLL fermions the reason of 
which will be clear as we proceed. However for simplicity, we consider $y_{l_{ij}}, \alpha_{l_{ij}}$ as diagonal 
having equal values $y_l$ and $\alpha_l$ respectively. Similarly $M_E$ and $M_{\chi}$ are taken as diagonal. 
We maintain the hierarchy (in case of three generations of VLLs) 
as $\Delta_{31} \gg \Delta_{32} \gg \Delta_{21}$ so that the DM phenomenology remains unaltered and effectively influenced by $m_1$ only ($i.e. ~ m_1$ only is involved in effective annihilation cross section of DM). Although the presence of several generations would lead to several mixing angles (involved through the mass matrix including the neutral components of all the VLL singlets and doublets) in general, we assume only one mixing, denoted by $\theta_l$, to be effective and ignore other mixings. Following our understanding of the DM 
analysis, we have fixed certain parameters at their benchmark values as: 
$\alpha_l = 0.01, \Delta_{21}$ = 50 GeV, $m_1$ = 850 GeV, $m_S$ = 840 GeV and  $\sin\theta_l$ = 0.1.
Later we will investigate the variation of $\alpha_l, \theta_l$ and its impact on the vacuum stability.

\begin{figure}[]
\centering
\subfigure[]{
\includegraphics[scale=0.395]{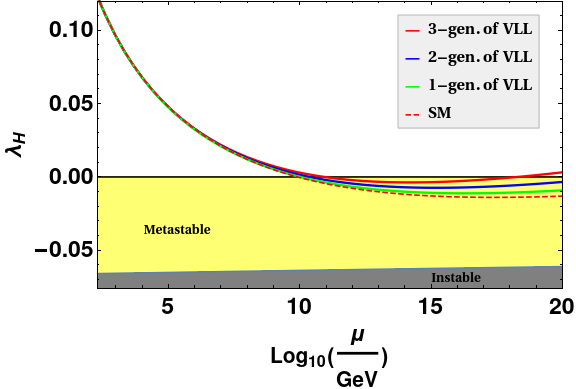}}
\subfigure[]{
\includegraphics[scale=0.395]{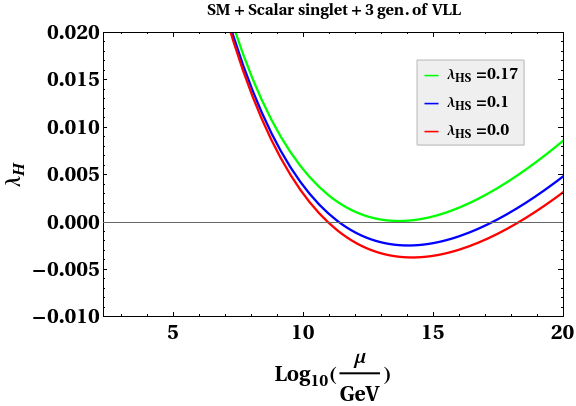}}
\caption{(a) Running of Higgs quartic coupling 
$\l_H$ against the scale $\mu$  by considering the presence of different generations of vector like leptons, while setting  $\l_{HS}$ = 0, (b) running of Higgs quartic coupling for different values of $\l_{HS}$ for the three generations of vector like leptons.}
\label{quartic-gen-imp}
\end{figure}

In order to understand the impact of VLLs on the running of the Higgs quartic coupling, let us first consider 
the left panel of Fig. \ref{quartic-gen-imp}.  In this plot, we provide running of the SM Higgs quartic coupling 
$\l_H$ against the scale $\mu$ considering the presence of different generations of VLL pairs 
of doublet and singlet, while presence of the scalar singlet DM is ignored by setting $\l_{HS}$ = 0.  It turns 
out that in contrary to the case of VLQs, presence of three generations of VLL doublet and singlet alone 
cannot keep $\l_{H}$ positive at all scale. As we find in the right panel of Fig. \ref{quartic-gen-imp}, a sizeable  
$\l_{HS} = 0.17$ along with three generations of VLLs (doublet and singlet) can keep the EW vacuum 
absolutely stable all the way till Planck scale (through the additional contribution in Eq. \eqref{run-LH}). 

In the right panel of Fig. \ref{quartic-gen-imp}, we notice that $\l_H$ becomes negative around $10^{11}$ GeV 
and again approaches positive before $M_{\rm Pl}$. Hence the situation with smaller $\l_{HS}$, say 0.1 with 3 generations of VLL, indicates existence of another deeper minimum and EW vacuum is found to be metastable. 
This is true even for 3 generations of VLL with $\l_{HS}$ =0. Note that it requires $\l_{HS}$ to be 0.17 in order to 
make the vacuum absolutely stable. So for the rest of the analysis, where we stress upon the absolute stability 
only, we consider $\l_{HS} = 0.2$ and set the DM mass at 840 GeV. On the other side, if we accept the 
metastability of the EW vacuum (with $\l_{HS} < 0.17$ having 3-generation VLLs) as a possibility, then a lighter 
DM mass (along with $m_1 \gtrsim m_S$) can also be realised in the present framework.

\begin{figure}[]
\centering
\subfigure[]{
\includegraphics[scale=0.39]{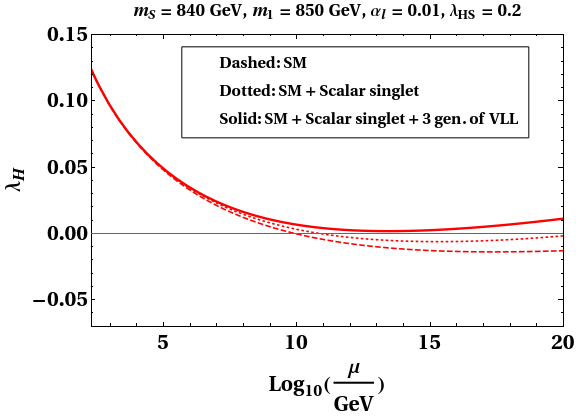}}
\subfigure[]{
\includegraphics[scale=0.38]{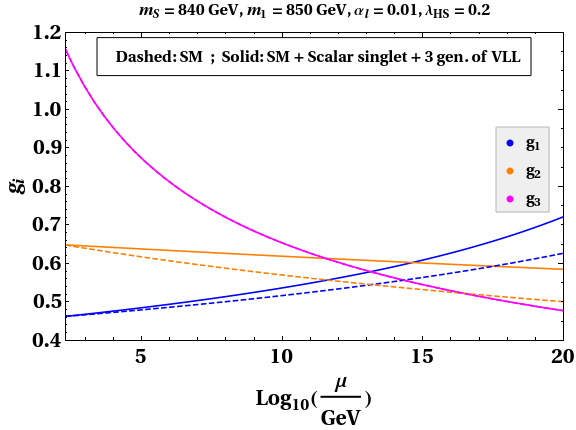}}
\caption{(a) Running of Higgs quartic coupling 
$\l_H$ against the scale $\mu$ for the SM, SM + scalar singlet and SM + scalar singlet + 3 generation of VLLs scenario  (b) running of  gauge couplings against the scale $\mu$ for the SM and SM + scalar singlet + 3 generation of VLLs scenario.}
\label{running1vll}
\end{figure}

We now focus on the details of the specific pattern of running of $\l_{H}$ in this scenario as compared to the 
VLQ case. In Fig. \ref{running1vll}, $\l_H$ running is shown in left panel for three different cases: the dashed red line 
correspond to the SM variation while dotted and solid lines stand for SM + scalar singlet DM (with $\l_{HS}$ 
= 0.2) and SM+scalar singlet DM + 3 generations of VLLs. In the right panel, running of the gauge couplings $g_{1,2,3}$ are shown. We notice that while $g_3$ remains unaffected, $g_1$ and $g_2$ undergo a positive 
shift (compared to the SM running) due to the presence of VLL. These shifts can be understood by looking 
at the additional contributions to the respective $\b$ functions as in Eq. \eqref{beta_g1g2} as 
\besub
\bea
\b{g_1}  &=&  \b_{g_1}^{\rm{SM}} +\b_{g_1}^{\rm{VLL(1)}}~ = ~ \b_{g_1}^{SM} + \frac{g_1^3}{16 \pi^2}\bigg{[}\frac{4}{5}(2n_2Y_{E}^2+n_1Y_{\chi}^2)\bigg{]},  \\
\b{g_2}  &=&  \b_{g_2}^{\rm{SM}} +\b_{g_2}^{\rm{VLL(1)}}~ = ~  \b_{g_2}^{SM} + \frac{g_2^3}{16 \pi^2}\bigg{(}\frac{2}{3}n_2\bigg{)},
\eea
\label{beta_g1g2_vll}
\eesub
\noindent where $n_2$ is the number of $SU(2)$ doublets, $n_1$ is the number of $SU(2)$ singlets, $Y_{E, \chi}$ are the hypercharges of $E$ and $\chi$ respectively\footnote{Although here we represent the one-loop beta functions for the BSM part for discussion purpose, we finally include the two-loop beta functions while studying the running.}. 
Though this helps in decreasing the $y_t$ value due to running, it is not adequate to keep $\l_{H}$ positive at any scale beyond $\Lambda^{\rm SM}_I$ even with SM+ three generations of VLLs. In lifting the $\l_{H}$ to positive 
side for any scale till $M_{\rm Pl}$ or to maintain the absolute stability, contributions of both $\l_{HS}$ (through the additional term in $\b_{\l_{H}}$ similar to Eq. \eqref{run-LH}) and 3 generations of VLL 
(due to the presence of $n_1, n_2$ in $\b_{g_1, g_2}$ via Eq. \eqref{beta_g1g2}) turn out to be important.

\subsection{Combined analysis of DM and EW vacuum stability}

\begin{figure}[]
\centering
\subfigure[]{
\includegraphics[scale=0.39]{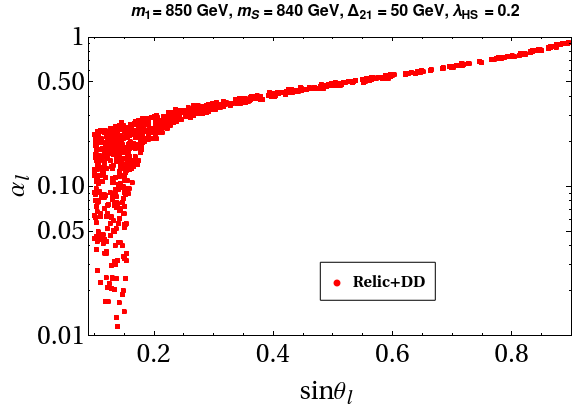}}
\subfigure[]{
\includegraphics[scale=0.39]{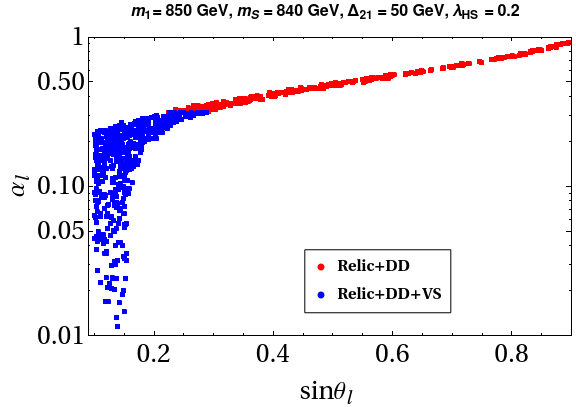}}
\caption{Contour plot showing allowed parameter space from (a) the DM constraints, (b) DM and the vacuum stability constraints in the $\a_l-\sin\theta_l$ plane.}
\label{alpha-theta}
\end{figure}

In this section, we consider the effect of varying the two parameters $\alpha_l$ and $\theta_l$. Therefore 
we perform a scan over a wide range of $\alpha_l$ and $\theta_l$ and represent the correlation plot obtained 
in the left panel of Fig. \ref{alpha-theta}. These points in red correspond to the correct DM relic density 
and simultaneously they satisfy the DD constraints. In generating the plot, we fix $m_S$ = 840 GeV, 
$m_1 =$ 850 GeV, $\Delta_{21} = 50$ GeV and $\l_{HS} = 0.2$. We find that for larger $\sin\theta_l$, 
only large values of $\alpha_l$ are allowed. Note that contrary to the VLQ case, $\alpha_l$ does not 
participate in the DD cross section as the associated interaction involves leptons only. Hence from the 
DD point of view, no restriction can be imposed on $\alpha_l$. For the criteria of satisfying relic density, as 
we have seen in DM sub-section that it is mainly due to the annihilation of VLLs via gauge mediation 
which contributes to the required DM annihilation cross section in small mixing angle limits. Now for large 
$\theta_l$, as we have already pointed out in subsection \ref{DM_VLL} in the context of Fig. \ref{VLL-relic-dd-plots}, a relatively 
large mixing angle $\sim 0.9$ (indicating $m_1$ is mostly made up of singlet VLL) is disallowed 
as it won't correspond to sufficient annihilation of this type (gauge mediated processes) required to 
satisfy the relic. However the situation changes when $\alpha_l$ is also large. This is due to the fact that 
co-annihilation processes of DM involving large $\alpha_l$ can provide the required relic. 

In the right panel of Fig. \ref{alpha-theta}, we introduce the constraints from vacuum stability. The blue 
portion now denotes the allowed region of parameter space in $\alpha_l$-$\theta_l$ plane, while the 
red portion is disallowed from vacuum stability constraints, in particular due to the violation of co-positivity 
criteria of Eq. \eqref{copos}. In this context, it can be noted that terms proportional to $\alpha^4$ present in 
$\b_{\l_{S}}$ (see appendix \ref{RGE_VLL} for its expression) makes $\l_{S}$ negative at some high scale. Hence 
applying the co-positivity conditions, the restricted parameter space for $\alpha_l-\theta_l$ is obtained.

\section{Discussion and Conclusion}
\label{conclude}

We have studied the possibility of sub-TeV scalar singlet dark matter along with electroweak vacuum stability by incorporating the presence of additional vector like fermions. While a pure scalar singlet DM extension of the standard model can satisfy the requirement of correct DM phenomenology along with EW vacuum stability only for DM mass above a TeV, introduction of additional vector like fermions is shown to bring it down to sub-TeV regime. Due to the presence of new co-annihilation channels between DM and vector like fermions (most importantly the annihilations of VLF), it is possible to satisfy DM relic criteria without getting into conflict with the direct detection data. The same additional fermions also play an instrumental role in keeping the Higgs quartic coupling positive 
at all scales up to $M_{\rm Pl}$ dominantly through their contributions to the RG evolution of SM gauge couplings. 

Similar to the extension with VLQ, the VLL extension also gives rise to additional annihilation and co-annihilation channels of DM, as can be seen from the interaction terms included in the Lagrangian. One advantage of this extension compared to the VLQ extension discussed above is that, it does not give rise to additional contribution to DM-nucleon scattering at radiative or one loop level . Also, the additional contribution to DM relic from Sommerfeld enhancement is absent in this scenario. While one can still have sub-TeV scalar singlet DM allowed from all relevant constraints, the criteria of absolute EW vacuum stability all the way till Planck scale requires at least three generations of such additional leptons along with a sizeable contribution from the DM portal coupling with the SM Higgs. This is precisely due to the difference in the way VLQs contribute to RG running of $SU(3)$ gauge coupling from the way VLLs contribute to $SU(2)$ (dominantly) gauge coupling where the latter do not have any additional colour degrees of freedom. 

\begin{figure}[H]
\centering
\includegraphics[scale=0.5]{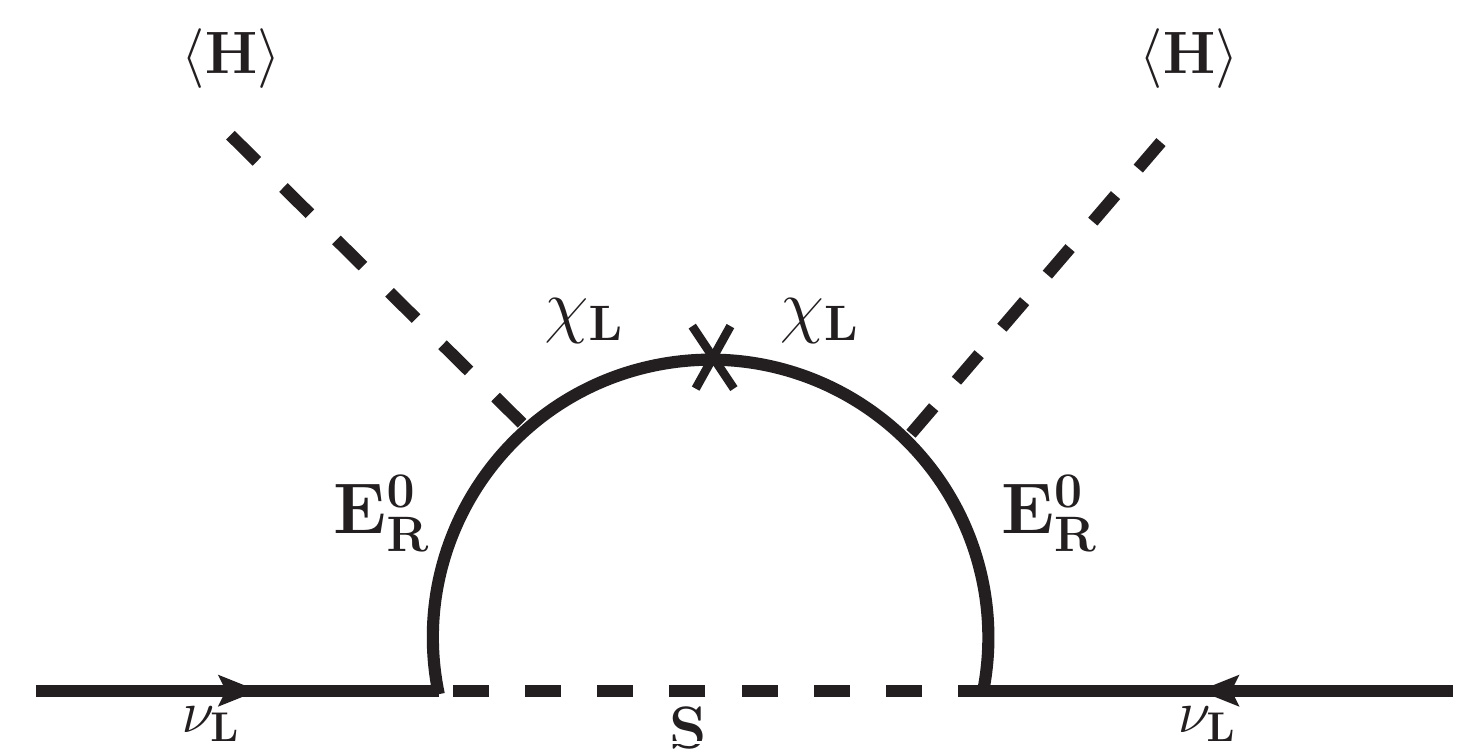}
\caption{$\nu-$mass generation at 1-loop}
\label{NM}
\end{figure}

While the construction with VLL may appear non-minimal compared to the VLQ extension, the additional three families of leptons can play a non-trivial role in generating light neutrino masses at one-loop level as shown in 
Fig. \ref{NM} \cite{Dias:2012xp,Klasen:2016vgl,May:2020bpo}. This however requires addition of Majorana mass terms for $\chi$ field, which are in fact allowed by the symmetry of the model. A similar construction has been exercised recently in \cite{Konar:2020wvl} in presence of Majorana masses 
for both $\chi_L$ and $\chi_R$. Since the very details of neutrino mass expression does not carry a direct connection with the DM and vacuum stability part of our scenario, we refrain from studying the details of 
neutrino mass here.  

We have studied the two extensions of the scalar singlet DM model by adopting a minimalistic approach. While our studies have been motivated primarily by the possibility of sub-TeV DM and EW vacuum stability, the scenarios can offer very rich phenomenology in terms of collider physics, indirect detection of dark matter, charged lepton flavour violation etc. While one can get additional contribution to dijet plus missing energy or dilepton plus missing energy signatures due to vector like quarks and leptons respectively, the same additional fermions can also boost DM annihilation to gamma rays, a tantalising indirect detection signature that has been searched for at several experiments. The leptonic extension of the model which also generates light neutrino masses at one loop, can in principle give rise to observable charged lepton flavour violation like $\mu \rightarrow e \gamma, \mu \rightarrow 3e$ etc, specially in the regime where singlet DM has order one Yukawa coupling with the additional fermions. Another possible direction is to consider a UV completion of this minimal model where $Z_2$ symmetry can be embedded in a gauge symmetry which also predict the scale of additional vector like fermions. We leave such aspects of our current scenario to future works.

\acknowledgments

DB acknowledges the support from Early Career Research Award from DST-SERB, Government of India (reference number: ECR/2017/001873). RR would like to thank Basabendu Barman, Abhijit Kumar Saha, Drona Vatsyayan and Lopamudra Mukherjee for various
useful discussions during the course of this work.  

\appendix
\label{app}

\section{1-loop $\b$-functions for VLQ scenario}
\label{RGE_VLQ}
Below we provide the 1-loop $\b$-functions for all the couplings involved in vector like quark scenario. While generating the $\b-$functions we have considered one complete family of VLQ  $i.e.$ one double and one singlet. In the expressions below color charge $N_c=3$ for all the VLQs, $n_1$ and $n_2$ represent number of $SU(2)$ singlet and doublet VLLs respectively while $n_3$ corresponds to the number of $SU(3)$ triplet vector like fields and finally $n_F$ represents the number of complete VLQ families coupled to the SM Higgs. Due to the involvment of one complete family of VLQs all $n_1$, $n_2$, $n_F$ are 1 whereas $n_3=3$. Here, $Y_{\mathcal{F}}=-\frac{1}{6}$ and $Y_{f{\prime}}=\frac{2}{3}$ are the hypercharges of $\mathcal{F}$ and $f{\prime}$ respectively. 

\subsubsection{SM Couplings}
{\allowdisplaybreaks  \begin{align} 
 \beta_{g_1}&= \b_{g_1}^{\rm{SM}} +\b_{g_1}^{\rm{VLQ}}~ = ~ \b_{g_1}^{\rm{SM}} + \frac{g_1^3}{16 \pi^2}\bigg{(}\frac{4}{5}N_c~(2n_2Y_{\mathcal{F}}^2+n_1Y_{f{\prime}}^2)\bigg{)} \nonumber  \\
& =\beta_{g_1}^{SM}+ 
\frac{1}{16\pi^2}\Big(\frac{6}{5} g_{1}^{3}\Big) \\  
\beta_{g_2}&= \b_{g_2}^{\rm{SM}} +\b_{g_2}^{\rm{VLQ}}~ = ~\b_{g_2}^{\rm{SM}}+ \frac{g_2^3}{16 \pi^2}\bigg{(}\frac{2}{3}N_c~n_2\bigg{)}\nonumber  \\
& = \beta_{g_2}^{SM}+\frac{1}{16\pi^2}( 
2 g_{2}^{3})  \\
\beta_{g_3} & = \b_{g_3}^{\rm{SM}} +\b_{g_3}^{\rm{VLQ}}~ = ~\b_{g_3}^{\rm{SM}} + \frac{g_3^3}{16 \pi^2}\bigg{(}\frac{2n_3}{3}\bigg{)}\nonumber  \\
& = \beta_{g_3}^{SM}+ 
\frac{1}{16\pi^2}(2 g_{3}^{3}) \\   
\beta_{\l_H} & = \b_{\l_H}^{\rm{SM}}+\b_{\l_H}^{\rm{S}} +\b_{\l_H}^{\rm{VLQ}}~ = ~\b_{\l_H}^{\rm{SM}}+\b_{\l_H}^{\rm{S}} + \frac{2n_F}{16 \pi^2}~(4N_cy_l\l_H-2N_cy^4)\nonumber  \\
& = \beta_{\l_H}^{SM} 
+\frac{1}{16\pi^2}\Big(\frac{1}{2} \lambda_{HS}^{2}-12 y^{4}+24 y^{2}\l_H\Big) \\    
\beta_{y_t} &= \b_{y_t}^{\rm{SM}} +\b_{y_t}^{\rm{VLQ}}~ = ~\b_{y_t}^{\rm{SM}} + \frac{n_Fy_t}{16 \pi^2}~(2N_cy^2)\nonumber  \\
& \beta_{y_t}^{SM}+ 
\frac{1}{16\pi^2}(6 y_ty^{2})   
\end{align}} 
\subsubsection{BSM couplings}
{\allowdisplaybreaks  \begin{align} 
\beta_{\l_S}^{(1)}&  = \frac{1}{16\pi^2}\Big( 
3 \Big(16 \lambda_S\alpha_{1}^2   -48 \alpha_{2}^4  + 4 \lambda_{HS}^{2}  + 8 \lambda_S \alpha_{2}^2  -96 \alpha_{1}^4  + \lambda_{S}^{2}\Big)\Big)\\ 
\beta_{\l_{HS}}^{(1)}&  =\frac{1}{16\pi^2}\Big(  
-\frac{9}{10} g_{1}^{2} \lambda_{HS} -\frac{9}{2} g_{2}^{2} \lambda_{HS} +4 \lambda_{HS}^{2} +\lambda_{HS} \lambda_S +12 \lambda_{HS} y^{2} +12 \lambda_{HS} \lambda \nonumber \\ 
 &+24 \Big(- y^{2}  + \lambda_{HS}\Big)\alpha_{1}^2 +24 y \alpha_{1}  y_{t}\alpha_{2} +24 y \alpha_{1}  y_{t}  \alpha_{2} -24 \alpha_{1}^2  y_{t}^2 +12 \lambda_{HS} \alpha_{2}^2 \nonumber \\ 
 &-24 y^{2} \alpha_{2}^2 -24 y_{t}^2  \alpha_{2}^2  +6 \lambda_{HS} y_t^2\Big) \\ 
\beta_{y}^{(1)}&  =  
   \frac{1}{16\pi^2}\Big(3 y y_t^2  -8 g_{3}^{2} y  + \frac{15}{2} y^{3}  -\frac{17}{20} g_{1}^{2} y  -\frac{9}{4} g_{2}^{2} y \Big) \\   
\beta_{\a_1}^{(1)}& = \frac{1}{16\pi^2}\Big( 
\frac{1}{10} \Big(5 \Big(-4 y y_{t} \alpha_{2} + y_{t}^2\alpha_{1}\Big) - \alpha_{1} \Big(-150 \alpha_{1}^2  + 45 g_{2}^{2}  -5 y^{2}  -60 \alpha_{2}^2 \nonumber \\ 
 & + 80 g_{3}^{2}  + g_{1}^{2}\Big)\Big)\Big)\\ 
 \beta_{\a_2}^{(1)}& = \frac{1}{16\pi^2}\Big( 
-4 y y_t  \alpha_{1}  + \alpha_{2} \Big(12 \alpha_{1}^2  -8 g_{3}^{2}  + 9 \alpha_{2}^2  -\frac{8}{5} g_{1}^{2}  + y^{2}\Big) + y_t^2\alpha_{2}\Big)  
\end{align}}

\section{1-loop $\b$-functions for VLL scenario (3 generations)}
\label{RGE_VLL}

Here, we provide the 1-loop $\b$-functions of all the relevant couplings by considering three complete generations of VLLs $i.e.$ three doublets and three singlets. In the expressions below color charge $N_c=1$ for all the VLLs, $n_1$, $n_2$, $n_F$ are 3 due to the presence of three complete generations of VLL families whereas $n_3=0$  and finally  $Y_{E}=-\frac{1}{2}$ and $Y_{\chi}=0$ are the hypercharges of $E$ and $\chi$. Here we have also assumed that $y_{l_{ij}}, \alpha_{l_{ij}}$ as diagonal having equal values $y_l$ and $\alpha_l$ respectively.    

\subsubsection{SM Couplings}
{\allowdisplaybreaks  \begin{align} 
 \beta_{g_1}&= \b_{g_1}^{\rm{SM}} +\b_{g_1}^{\rm{VLL}}~ = ~ \b_{g_1}^{\rm{SM}} + \frac{g_1^3}{16 \pi^2}\bigg{(}\frac{4}{5}N_c~(2n_2Y_{E}^2+n_1Y_{\chi}^2)\bigg{)} \nonumber  \\
& =\beta_{g_1}^{SM}+ 
\frac{1}{16\pi^2}\bigg{(}\frac{6}{5} g_{1}^{3}\bigg{)} \\  
\beta_{g_2}&= \b_{g_2}^{\rm{SM}} +\b_{g_2}^{\rm{VLL}}~ = ~\b_{g_2}^{\rm{SM}}+ \frac{g_2^3}{16 \pi^2}\bigg{(}\frac{2}{3}N_c~n_2\bigg{)}\nonumber  \\
& =\beta_{g_2}^{SM}+\frac{1}{16\pi^2}(2 g_{2}^{3})  \\
\beta_{g_3} & = \b_{g_3}^{\rm{SM}} +\b_{g_3}^{\rm{VLL}}~ = ~\b_{g_3}^{\rm{SM}} + \frac{g_3^3}{16 \pi^2}\bigg{(}\frac{2n_3}{3}\bigg{)}\nonumber  \\
& = \beta_{g_3}^{SM} 
 \\   
\beta_{\l_H} & = \b_{\l_H}^{\rm{SM}}+\b_{\l_H}^{\rm{S}} +\b_{\l_H}^{\rm{VLL}}~ = ~\b_{\l_H}^{\rm{SM}}+\b_{\l_H}^{\rm{S}} + \frac{2n_F}{16 \pi^2}~(4N_cy_l\l_H-2N_cy_l^4)\nonumber  \\
& =\beta_{\l_H}^{SM} +\frac{1}{16\pi^2}\bigg{(}\frac{1}{2} \lambda_{HS}^{2} +24 \lambda_{H} y_l^2 -12 y_l^4\bigg{)} \\    
\beta_{y_t} &= \b_{y_t}^{\rm{SM}} +\b_{y_t}^{\rm{VLL}}~ = ~\b_{y_t}^{\rm{SM}} + \frac{n_Fy_t}{16 \pi^2}~(2N_cy_l^2)\nonumber  \\
& \beta_{y_t}^{SM}+ 
\frac{1}{16\pi^2}(6y_t y_l^2)
\end{align}} 

\subsubsection{BSM couplings}
{\allowdisplaybreaks  \begin{align} 
\beta_{\l_S}&  = \frac{1}{16\pi^2}\Big( 
48 \lambda_S \a_l-288 \a_l^4 + 3 \Big(4 \lambda_{HS}^{2}  + \lambda_{S}^{2}\Big)\Big)\\ 
\beta_{\l_{HS}}&  =\frac{1}{16\pi^2}\Big(  
-\frac{9}{10} g_{1}^{2} \lambda_{HS} -\frac{9}{2} g_{2}^{2} \lambda_{HS} +4 \lambda_{HS}^{2} +\lambda_{HS} \lambda_S +12 \lambda_{HS} \lambda +12 \lambda_{HS}y_l^2 \nonumber \\ 
 &+ 24 \lambda_{HS} \a_l^2 +6 \lambda_{HS}y_t^2-24 y_l^2\a_l^2\Big) \\ 
\beta_{y_l}&  = \frac{1}{16\pi^2}\Big( 
   \frac{15}{2} y_l^3 + y_l \Big( 3 y_t^2  -\frac{9}{20} g_{1}^{2}  -\frac{9}{4} g_{2}^{2}  \Big)\Big)\\   
\beta_{\a_1}& = \frac{1}{16\pi^2}\Big( 
\frac{1}{10} \Big(5y_l^2\a_l + 150\a_{l}^3  -9\a_l \Big(5 g_{2}^{2}  + g_{1}^{2}\Big)\Big)\Big)
\end{align}}

\bibliographystyle{apsrev}
\bibliography{ref.bib}

\end{document}